\documentclass[preprint,5p]{elsarticle}
\usepackage{graphicx} 
\usepackage{url}
\usepackage{amsmath}
\usepackage{listings}
\usepackage{multicol}
\usepackage{subcaption}
\usepackage{xcolor} 

\newcommand{\Tab}{Table}
\newcommand{\Fig}{Fig.}
\newcommand{\Figs}{Figs.}

\usepackage{physics}

\title{PCMS: Parallel Coupler For Multimodel Simulations}

\journal{Computer Physics Communications}

\begin{document}
\begin{frontmatter}
\author[a,b]{Jacob S. Merson\corref{author}}
\author[b]{Cameron W. Smith}
\author[a,b]{Mark S. Shephard}
\author[a]{Fuad Hasan}
\author[a]{Abhiyan Paudel}
\author[b]{Angel Castillo-Crooke}
\author[b]{Joyal Mathew}
\author[b]{Mohammad Elahi}
\cortext[author]{Corresponding author\\\textit{E-mail address:} mersoj2@rpi.edu}
\address[a]{Department of Mechanical, Aerospace, and Nuclear Engineering, Rensselaer Polytechnic Institute, 110 8th St. Troy, NY 12180}
\address[b]{Scientific Computation Research Center (SCOREC), Rensselaer Polytechnic Institute, 110 8th St. Troy, NY 12180}

\begin{abstract}
This paper presents the Parallel Coupler for Multimodel Simulations (PCMS), a new GPU accelerated generalized coupling framework for coupling simulation codes on leadership class supercomputers. PCMS includes distributed control and field mapping methods for up to five dimensions. For field mapping PCMS can utilize discretization and field information to accommodate physics constraints. PCMS is demonstrated with a coupling of the gyrokinetic microturbulence code XGC with a Monte Carlo neutral transport code DEGAS2 and with a 5D distribution function coupling of an energetic particle transport code (GNET) to a gyrokinetic microturbulence code (GTC). Weak scaling is also demonstrated on up to 2,080 GPUs of Frontier with a weak scaling efficiency of 85\%.
\end{abstract}

\end{frontmatter}

\section*{Program Summary}
\begin{small}
\noindent
{\em Program title:} Parallel Coupler for Multimodel Simulations                                          \\
{\em CPC library link to program files:} (to be added by Technical Editor) \\
{\em Developer's repository link:} https://github.com/SCOREC/pcms \\
{\em Licensing provisions(please choose one):} BSD 3-clause  \\
{\em Programming language:} C++                                   \\
{\em Supplementary material:} None                                \\
{\em Nature of problem:}
Multiscale and multiphysics modeling requires coupling of disparate simulation methodologies. Strategies that require rewriting numerics in a unified framework can yield good performance and stability, however, it is not practical for codes that require high-dimensional data, physics-based coordinate systems, the use of exascale computing resources, and other specialized numerical methods. Additionally, unified frameworks cannot easily take advantage of existing simulation infrastructure or independently developed AI models as is needed for digital twins. \\
{\em Solution method:}
PCMS provides infrastructure to couple multiple simulation codes at scale on exascale systems. It uses scalable methodologies for transferring data between distributed applications. PCMS takes a hierarchical approach to field transfer using as much information as possible, such as mesh topology, to accommodate conservation and other physical constraints. In the context of limited information, alternative strategies, such as radial basis functions, are used to support field transfer. PCMS has been demonstrated on coupled fusion simulations that make use of volume coupling of high-dimensional data and complex coordinate systems.
\\
  
\end{small}

\section{Introduction}
Simulation tools that can handle multiple types of physics are widely deployed in engineering practice. The increasing accessibility of exascale computers has similarly driven the advancement of highly specialized codes that can model complex devices such as fusion reactors and nanoscale microelectronic devices.

Effectively modeling these components and integrating them with machine learning and AI methods is difficult to achieve through monolithic modeling frameworks, where a consistent set of governing equations is solved together in a tightly coupled fashion. Instead, we must rely on a hierarchy of models that can be built upon existing simulation tools. This constraint comes from two sources. First, the computational cost of a monolithic approach would far exceed that of today's exascale supercomputers. The second, and more onerous, is that developing each independent code has required decades of multi-disciplinary effort across physics, math, and computer science. It is hard to overstate how difficult replicating this effort in a monolithic fashion that requires shared terminologies, notations, and conventions would be. 

Given the complexity of a monolithic approach, a number of tools have been developed to support field transfer and coupling between existing simulation codes \cite{bungartzPlugandplayCouplingApproach2015,chourdakisPreCICEV2Sustainable2022,gastonMOOSEParallelComputational2009,slatteryDataTransferKit2013}. preCICE focuses on surface coupling and was originally developed for fluid-structure interactions and conjugate heat transfer simulations \cite{bungartzPlugandplayCouplingApproach2015,chourdakisPreCICEV2Sustainable2022}. It uses radial basis functions for field transfer operations and largely treats meshes as point clouds \cite{bungartzPlugandplayCouplingApproach2015,chourdakisPreCICEV2Sustainable2022}. Moose is a monolithic multiphysics framework that is widely used in nuclear engineering and is gaining visibility for fusion reactor simulations \cite{gastonMOOSEParallelComputational2009}. The Data Transfer Toolkit (DTK) provides a wide range of solution transfer operators, but has not been updated to work with GPUs and has been archived on github as of 2023 \cite{slatteryDataTransferKit2013,ORNLCEESDataTransferKit2025}. Portage is a modern effort being developed at Los Alamos National Laboratory that can support GPU accelerated point-localization and CPU-based mesh intersection methods \cite{noauthor_laristraportage_2025}.

Although these currently available coupling tools could address a number of the critical coupling operations, none of them were well suited to support the full range of needs for doing volume coupling of massively parallel fusion codes. Coupling fusion codes has additional complexities originating from the use of non-standard discretization methods, physics-based coordinate systems, high-dimensional data (up to 6D), and the use of programming paradigms suitable for exascale systems. Many of these codes employ unstructured meshes over complex domains \cite{Shephard_2024}.

A brief review of multiscale modeling methods and how they are applied in fusion simulations is provided to motivate the requirements and constraints that have been used in the development of PCMS.

Multiscale modeling can largely be broken into two categories. Models that link information across discrete scales, typically called hierarchical or upscaling multiscale methods, and those that directly resolve fine-scale models and interface fine-scale model information directly with coarse-scale models, typically called concurrent or resolved-scale multiscale methods \cite{fishMesoscopicMultiscaleModelling2021,tadmorModelingMaterialsContinuum2011}. PCMS, addresses the needs of resolved-scale multiscale methods.

Resolved-scale methods are often described as either multiscale or multiphysics methods depending on the details of the physical models being linked \cite{keyesMultiphysicsSimulationsChallenges2013,tadmorModelingMaterialsContinuum2011,fishMesoscopicMultiscaleModelling2021}. The most common resolved-scale multiscale method uses domain decomposition where subscale models are used in regions that require fine-scale details such as at localization sites like crack tips or in nanoindentation. Significant literature was developed in the late 1990's and early 2000's to link atomistic to continuum simulations in a resolved-scale fashion. Examples include the quasicontinuum method \cite{tadmorQuasicontinuumAnalysisDefects1996}, bridging scale method \cite{wagnerCouplingAtomisticContinuum2003}, concurrent AtC method \cite{fishConcurrentAtCCoupling2007}, and others \cite{xuConcurrentCouplingAtomistic2010,millerUnifiedFrameworkPerformance2009,parksConnectingAtomistictoContinuumCoupling2008}. The use of domain decomposed resolved-scale multiscale methods have recently been extended to use machine learned surrogates such as Deep Operator Networks (DeepONets) in the fine-scale domain to avoid remeshing \cite{yinInterfacingFiniteElements2022}.

Resolved-scale multiphysics simulations employ a decomposition strategy, utilizing different physical models and numerical schemes in each portion of the domain. Examples of this include linking inexpensive core plasma solvers to expensive edge plasma solvers (i.e., core-edge integration) \cite{dominskiSpatialCouplingGyrokinetic2021,merloFirstCoupledGENE2021}  and linking Monte Carlo computed neutron heating to continuum-based thermomechanical and CFD models \cite{novakMonteCarloMultiphysics2024}, and numerous other applications \cite{keyesMultiphysicsSimulationsChallenges2013}.

The Parallel Coupler for Multimodel Simulations (PCMS) is being developed to meet the coupling needs of such simulation codes. Aspects of PCMS critical to meet the desired simulation code coupling requirements are:
\begin{itemize}
    \item Integration of coordinate and data transformations in multidimensional domains to support unique and non-linear forms of coordinate systems and field data in the frequency domain requiring extra processing.
    \item Distributed control and transfer algorithms that account for details of field representations over multiple dimensions by allowing the use, or implementation of, native interrogation methods (such as interpolation) wherever possible.
    \item A design that avoids intrusive simulation code changes and makes use of existing data structures and methods.
    \item Employs a high level, discretization independent, specification of the simulation problem to support fully general non-manifold geometric domains\cite{weiler1985edge,beall2004comparison}.  
\end{itemize}

PCMS is designed to streamline the evaluation and usage of coupling schemes by applied mathematicians and physicists, eliminating the need for intrusive changes to coupled physics codes, while affording the ability to satisfy field transfer accuracy and conservation requirements.

This paper is organized as follows. The overall approach and high-level software components are described in section~\ref{sec:coupling-method}. In sections~\ref{sec:field-transfer}--\ref{sec:parallel-control}, the major components, field transfer and distributed control, are covered in detail. A number of fusion-relevant examples are provided in section~\ref{sec:examples}. Performance and scaling results are presented in section~\ref{sec:performance}. Section \ref{sec:integration} discusses integration of PCMS with workflow management tools and fusion-specific interfaces. Section~\ref{sec:future-work} provides a brief conclusions and discussion of future work.

\section{Code Coupling Software Methodology} \label{sec:coupling-method}

PCMS is split into two components: the ``coupler'', or ``server'', and a thin shim layer that is called the ``application'' or ``client''. This approach leads to the ability to couple disparate applications with a minimal amount of modifications to application build systems and data structures and algorithms. The coupler is designed to utilize an intermediate representation that can handle the full range of geometric, field, and coordinate system complexity while the shim layer is designed to be as lightweight as possible to enable the minimum set of operations needed for coupling.

The fusion community has multiple codes and reduced order models that can be used in each part of the reactor volume with variations on the physical simplifications or solution methodologies. It is desired to be able to swap these models to investigate the impact of physical and numerical modeling choices. When coupling is performed naively, it suffers quadratic scaling with the number of codes that need to be coupled. Using an intermediate representation ameliorates this challenge, requiring a single coupling implementation to the intermediate representation. PCMS interacts with the intermediate representation through the definition of compile-time interfaces.

The compile time interaction is done through an adapter base class that handles serialization, deserialization, setting up the map to the coupler partition, and defining a set of global IDs for the degrees of freedom.



\begin{figure}
    \centering
    \includegraphics[width=0.5\textwidth]{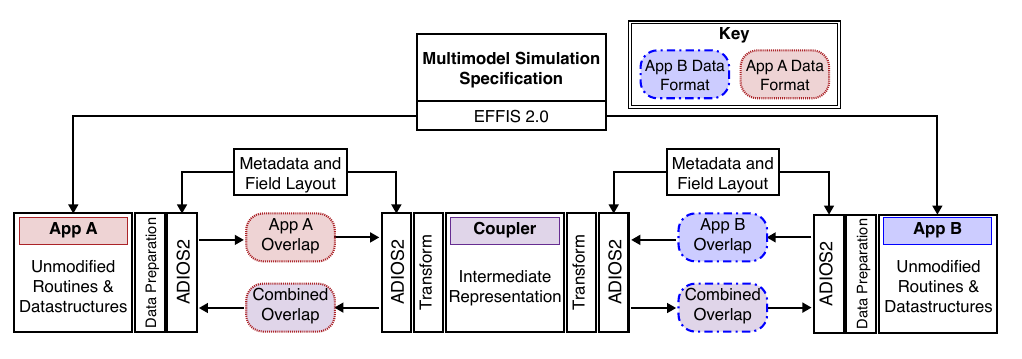}
    \caption{Overview of the PCMS components and operations reprinted from \cite{Shephard_2024}. Each application uses native datastructures and the coupler uses an intermediate representations. Inter application communication is performed with ADIOS2 and EFFIS provides job control.}
    \label{fig:PCMS-components}
\end{figure}

The coupler communicates with each application through ADIOS2's \cite{godoyADIOSAdaptableInput2020} in-memory or file-based IO Engines. The coupler can be run as a separate application or as part of the analysis code to be coupled. Although using the coupler as an independent application requires additional data transfers, we have found that in many cases the coupler and each application may not be able to be able to utilize the same compilers, flags, etc preventing directly linking the coupler with the application. This is largely due to the complex and rapidly evolving toolchain and 3rd party version incompatibilities that are common on leadership class supercomputers. Typically, a thin shim class is used in each application that describes a hierarchy of metadata related to the coupling operation. These wrappers support the full range of PCMS distributed control and field transfer operations. However, PCMS can also operate with a limited subset of information in an ADIOS2 stream written directly by the application. \Fig~\ref{fig:PCMS-components} shows the interactions of these components including model specification and job launch utilities performed by EFFIS \cite{suchyta2022exascale}. Alternative choices for workflow management are available and described in section~\ref{sec:integration}.

PCMS ultimately performs three high-level tasks.
\begin{enumerate}
    \item field transfer operations
    \item distributed control and communication of fields
    \item coordinate transformations 
\end{enumerate}

\section{Design of Field Transfer Functionality} \label{sec:field-transfer}
\subsection{Coupling Modes}
PCMS aims to support coupling in a wide range of scenarios including situations when the the coupler does not have intrinsic knowledge of the target applications coordinate systems, shape functions etc. Here, we define the set of basic coupling modes:
\begin{description}
    \item[intrinsic] the coupler understands enough information to perform field transfer operations locally. When using conservative mesh-based field transfer, this includes coordinate systems, shape functions, and discretization.
    \item[extrinsic] the coupler uses remote field evaluations to perform field transfer. In this mode, the coupler only needs to understand the discretization and relies on the application for field evaluations and shape function evaluation.
\end{description}

Conservative field transfer methods are supported in each of these modes. For extrinsic coupling, point evaluations, which may be quadrature points, are batched and sent as a request to the application which locally performs evaluations and sends batched data back to the coupler. The coupler performs integration, localization, and linear algebra solutions as detailed below.

\subsection{Field Transfer Operations}
Field transfer, the mapping of a field described on one domain and discretization onto a new domain or discretization, plays an important role in resolved-scale multiscale and multiphysics simulations. Examples of field transfer include mapping velocities and forces in fluid structure interaction simulations \cite{mehlParallelCouplingNumerics2016,parkerComponentbasedParallelInfrastructure2006}, and mapping discrete momentum to continuum stresses in atomistic-continuum coupling \cite{wagnerCouplingAtomisticContinuum2003}. In fusion simulations, field mapping has many uses such as spatial volume coupling of core and edge plasma simulations \cite{dominskiSpatialCouplingGyrokinetic2021,chengSpatialCoreedgeCoupling2020,merloFirstCoupledGENE2021}, linking neutral particle and neutral beam simulations to plasma simulations (Fig. \ref{fig:5d-distribution-function}, Fig. \ref{fig:xgc-degas2-conservative-coupling}), and enabling the connection of non-conformal meshes that align with the magnetic field lines \cite{joFieldalignedGyrokineticSolver2025}.

Various field transfer methodologies exist and the choice is largely driven by cost, accuracy, and physics constraints such as the need for conservation of the underlying fields. Broadly speaking, these methodologies either use a direct evaluation of the source field at the target fields's node points (interpolation/extrapolation), or solve a secondary problem to minimize some residual measure of the error in the target field (projection). Projection based methods tend to be more expensive, but make it easier to bring in physics constraints. Some authors have managed to develop conservative interpolation methods \cite{alauzetParallelMatrixfreeConservative2016}.

The field evaluation methods can further be categorized into mesh-based methods, or pointwise methods. Mesh-based methods use the underlying structure of the discretization whereas the pointwise methods consider the nodal points of the source data as a point cloud and impose shape functions that are independent of the discretization that is used in the solution procedures. The most common pointwise evaluation method is based on radial basis functions (RBF) \cite{slatteryMeshfreeDataTransfer2016}. This approach is convenient in the context of coupling as it doesn't require the coupler to handle the complexities of the mesh structure. This approach has found wide success in other coupling libraries and is used to get conservative particle to cell transfers in Particle-In-Cell (PIC) codes \cite{bungartzPreCICEFullyParallel2016,mollenImplementationHigherorderVelocity2021}. \Citeauthor{slatteryMeshfreeDataTransfer2016} compared radial basis functions and mesh intersection methods and concluded that the accuracy and conservative properties of RBF and mesh-based methods are dependent on the geometry and field structure \cite{slatteryMeshfreeDataTransfer2016}.

Most mesh-based conservative field transfer methods utilize the approach presented in \cite{jiaoCommonrefinementbasedDataTransfer2004}. In this method the fields are conservatively transferred by creating an intersection mesh made from elements where the basis functions of both the source and target fields are continuous and smooth affording the use of  standard quadrature rules \cite{blanchardHighOrderAccurate2016,jaimanConservativeLoadTransfer2006,menonConservativeInterpolationUnstructured2011,farrellConservativeInterpolationVolume2011,farrellConservativeInterpolationUnstructured2009}.

PCMS supports a range of field transfer mechanisms ranging from those that maximize flexibility and require limited information about the geometry and discretizations, to those that require a full understanding of the discretizations but can more faithfully preserve physics constraints. All operations can be performed in parallel on CPUs or GPUs by making use of the Kokkos~\cite{kokkos2014} performance potability library.

\subsubsection{Local Weighted Polynomial Fitting}\label{sec:local-weighted-polynomial}
When no discretization information is known, PCMS utilizes a local weighted polynomial fitting method. The weighted minimization problem is formulated as
\begin{equation}
    \underset{c}{\text{min}} ||\vb*{\phi} \cdot (\vb{A}\vb{c} - \vb{b})||_2^2 + \lambda||\vb{c}||_2^2,
\end{equation}
where \(\vb{A}\) is the Vandermonde matrix, \(\vb*{\phi}\) is the diagonal weight matrix, \(\lambda\) is the regularization parameter, \(\vb{b}\) is the vector of source field values. To regularize the system we make use of Tikhonov regularization (ridge regression).

For each target point \(t_i\), a set of source points 
 \[\vb{s}^i = \{s_j | \phi^i_j=f(s_j, t_i, \alpha)>0\} \forall j \in n_s\]
is determined through a selection function \(f\) where \(s_j\) is the $j^\text{th}$ source point, \(\phi\) is the weighting function and \(\alpha\) represents an arbitrary number of additional parameters.

\begin{figure}
    \centering
    \includegraphics[width=\linewidth]{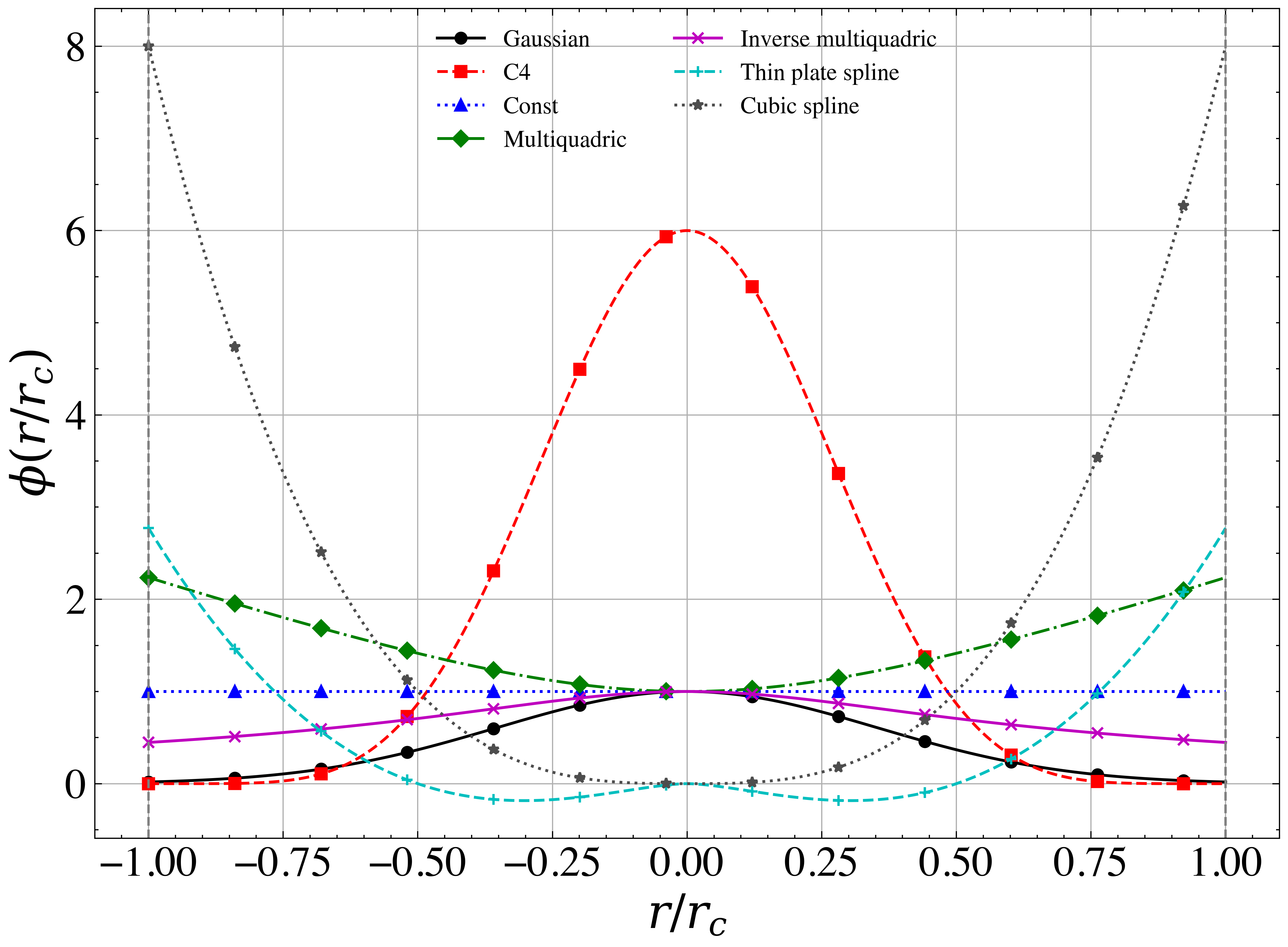}
    \caption{Radial basis functions implemented in PCMS (a = 2).}
    \label{fig:radial-basis-functions}
\end{figure}

A common choice for the selection function is to use the radius of the source point from the target point. This selection function is often combined with a radial basis function which describes the weight as a function of the distance away from the target point normalized by the cutoff distance ($r/r_c$). The definitions of the RBF that have been implemented in PCMS are listed in \Tab~\ref{tab:supported-basis-functions} and plotted in \Fig~\ref{fig:radial-basis-functions}.

If the cutoff distance is directly specified by the user, an error is thrown if the resulting minimization problem will be underdetermined for the specified polynomial order. Alternatively, the user can specify a minimum number of source points they want to use and an initial radius. The cutoff radius will be iteratively increased until every target point meets the user constraint.

Another common selection function selects the degrees of freedom contained in layers of elements that form patches in the source mesh around target point of interest. This is typically used with a constant weight of one which is equivalent to the patch recovery method that's used in superconvergent patch recovery (SPR) \cite{zienkiewiczSuperconvergentPatchRecovery1992} and Zienkiewicz and Zhu (ZZ) error estimators \cite{zienkiewiczSuperconvergentPatchRecovery1992,zienkiewiczSuperconvergentPatchRecovery1992b}.

\begin{table*}
\centering
\caption{Supported radial basis functions in PCMS for point-based field transfer.}
\label{tab:supported-basis-functions}
\begin{tabular}{|l|l|}
\hline
\textbf{Basis Function} & \textbf{Basis Function} \\ \hline
Gaussian                  & $\phi = \begin{cases} 0 & \text{for } r>r_c\\ \exp{-(ar/r_c)^2} & \text{for } r<r_c \end{cases}$                       \\ \hline
C4                        &  $\phi = \begin{cases} 0 & \text{for } r>r_c\\ \left[5(r/r_c)^5 + 30(r/r_c)^4 +72 (r/r_c)^3+82(r/r_c)^2+36(r/r_c)+6\right] (1-r/r_c)^6 & \text{for } r<r_c \end{cases}$                       \\ \hline
Const                     &  $\phi = \begin{cases} 0 & \text{for } r>r_c\\ 1 & \text{for } r<r_c \end{cases}$                       \\ \hline
No\_Op (identity)         &   $\phi=1$                      \\ \hline
Multiquadric              &  $\phi = \begin{cases} 0 & \text{for } r>r_c\\ \sqrt{1 + (ar/r_c)^2} & \text{for } r<r_c \end{cases}$  \\ \hline 
Inverse multiquadric           &  $\phi = \begin{cases} 0 & \text{for } r>r_c\\ 1/\sqrt{1 + (ar/r_c)^2} & \text{for } r<r_c \end{cases}$  \\ \hline
Thin plate spline           &  $\phi = \begin{cases} 0 & \text{for } r>r_c\\ (ar/r_c)^2\ln{(ar/r_c)} & \text{for } r<r_c \end{cases}$  \\ \hline
Cubic spline           &  $\phi = \begin{cases} 0 & \text{for } r>r_c\\ (ar/r_c)^3 & \text{for } r<r_c \end{cases}$  \\ \hline

\end{tabular}
\end{table*}

Although local weighted polynomial fitting provides advantages for flexibility and parallelizability this locality prevents the ability to globally conserve quantities. Many physical processes have quantities such as energy or mass that are conserved. Therefore, it is useful to have field transfer methods that can map field data in a conservative way. In the context of this work, we define two error measures to evaluate field value accuracy and the preservation of integral quantities.

The accuracy of the field transfer operation is defined as
\begin{equation}
    \text{Accuracy Error} = \frac{|| f_s(s_i)-f_s^*(s_i)  || }{||f_s^*(s_i)||}
\end{equation}
where \(f_s\) is the value on the source field after a mapping has made from the source to the target back to the source, and \(f^*_s\) is the ground truth which may be taken as the analytic function evaluated on the source mesh, or field values evaluated on the source mesh before any mapping has been performed.

The error of conservation is similarly defined as
\begin{equation}
    \text{Conservation Error} = \frac{|| \int f_s(s_i)-f_s^*(s_i) \dd{\Omega} || }{||\int f_s^*(s_i) \dd{\Omega}||}
\end{equation}
where, the conservation is evaluated with respect to an integral.

\begin{figure}
    \centering
    \begin{subfigure}[t]{0.48\linewidth}
        \centering
        \includegraphics[width=\linewidth]{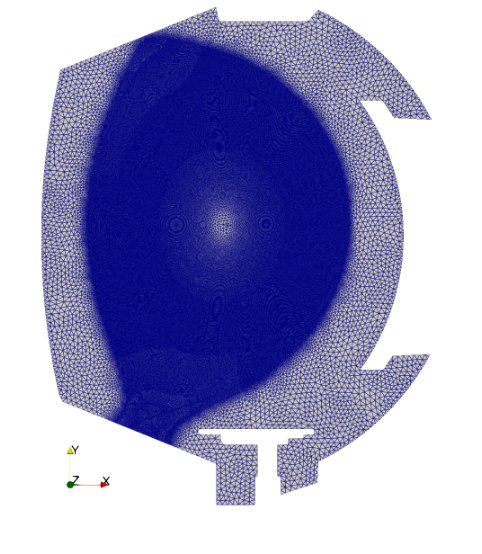}
        \caption{Field-following XGC meshes constructed using the TOMMS mesh generator \cite{riaz_automated_2024}. The mesh has 611,359 elements and 306,992 vertices.}
    \end{subfigure}
    ~
    \begin{subfigure}[t]{0.48\linewidth}
        \centering
        \includegraphics[width=\linewidth]{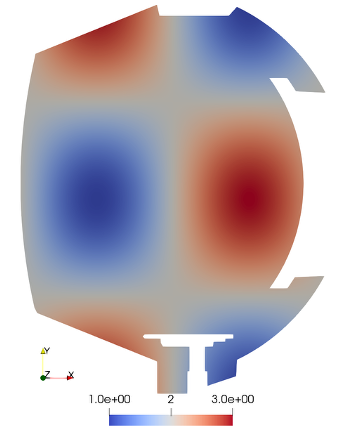}
        \caption{Analytic field (equation~\ref{eq:sin-cos-function}) used to measure the accuracy and conservation properties of RBF reconstruction on the field-following mesh.}
    \end{subfigure}
    \caption{Example mesh and field of WEST tokamak geometry obtained courtesy of Davide Curreli.}
    \label{fig:lcpp-meshes}
\end{figure}

\begin{figure}
    \centering
    \begin{subfigure}[t]{0.48\linewidth}
    \centering
    \includegraphics[width=\linewidth]{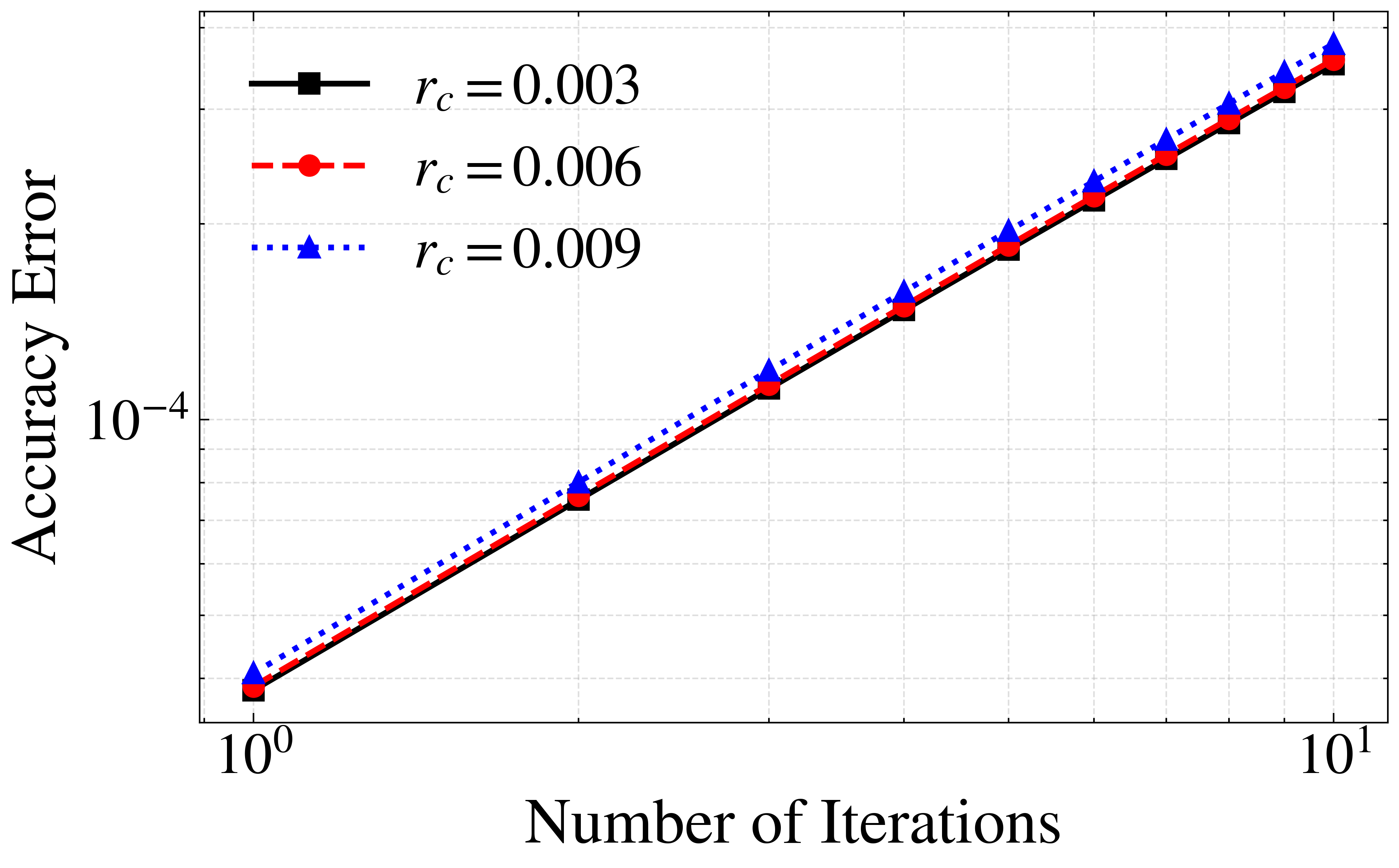}
    \caption{Accuracy}
    \end{subfigure}
    ~
    \begin{subfigure}[t]{0.48\linewidth}
    \centering
    \includegraphics[width=\linewidth]{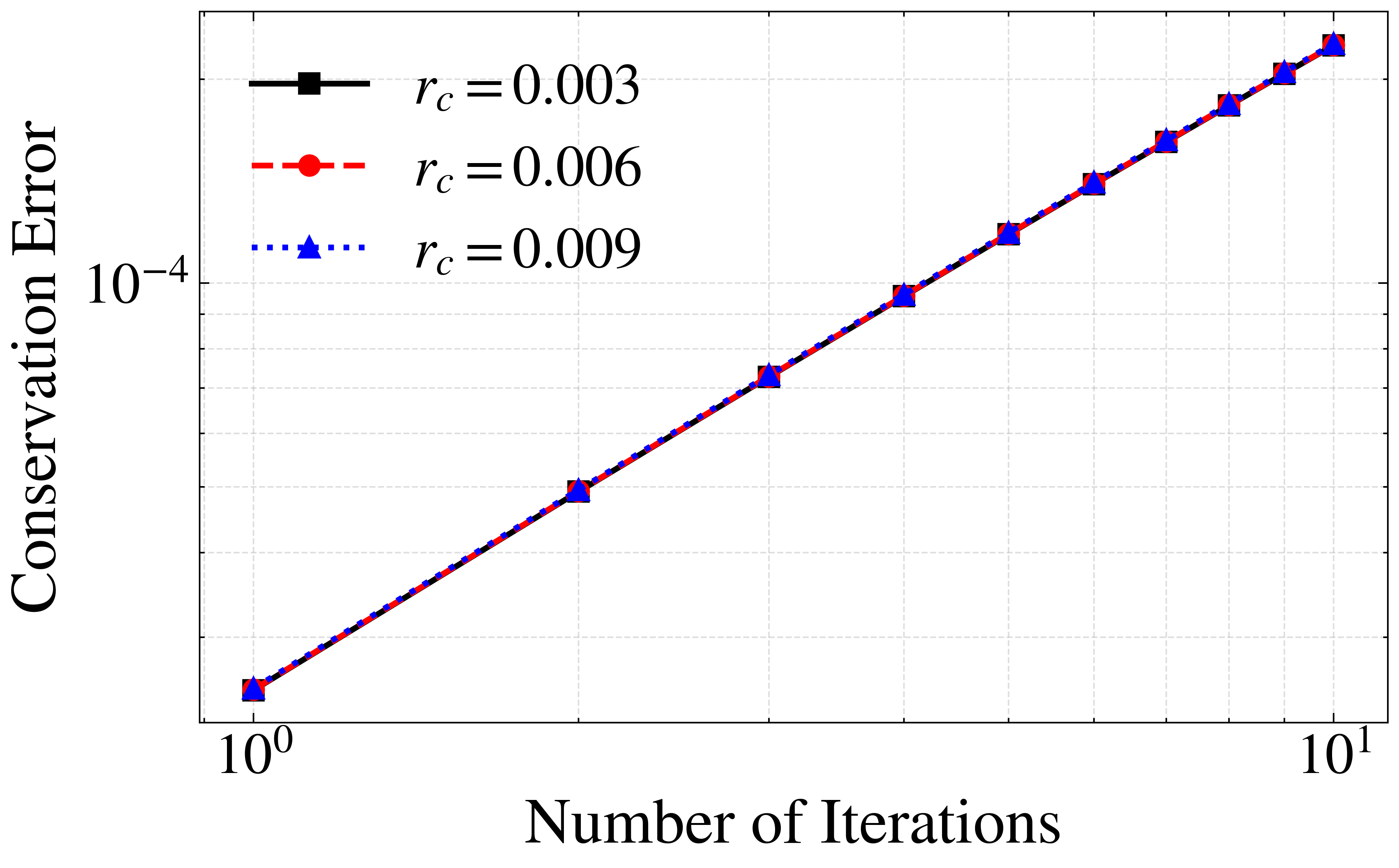}
    \caption{Conservation}
    \end{subfigure}
    \caption{Error of the linear field recovery when mapping from element centroids to vertices using C4 RBF.}
    \label{fig:linear-sincos}
\end{figure}
\begin{figure}
    \centering
    \begin{subfigure}[t]{0.48\linewidth}
    \centering
    \includegraphics[width=\linewidth]{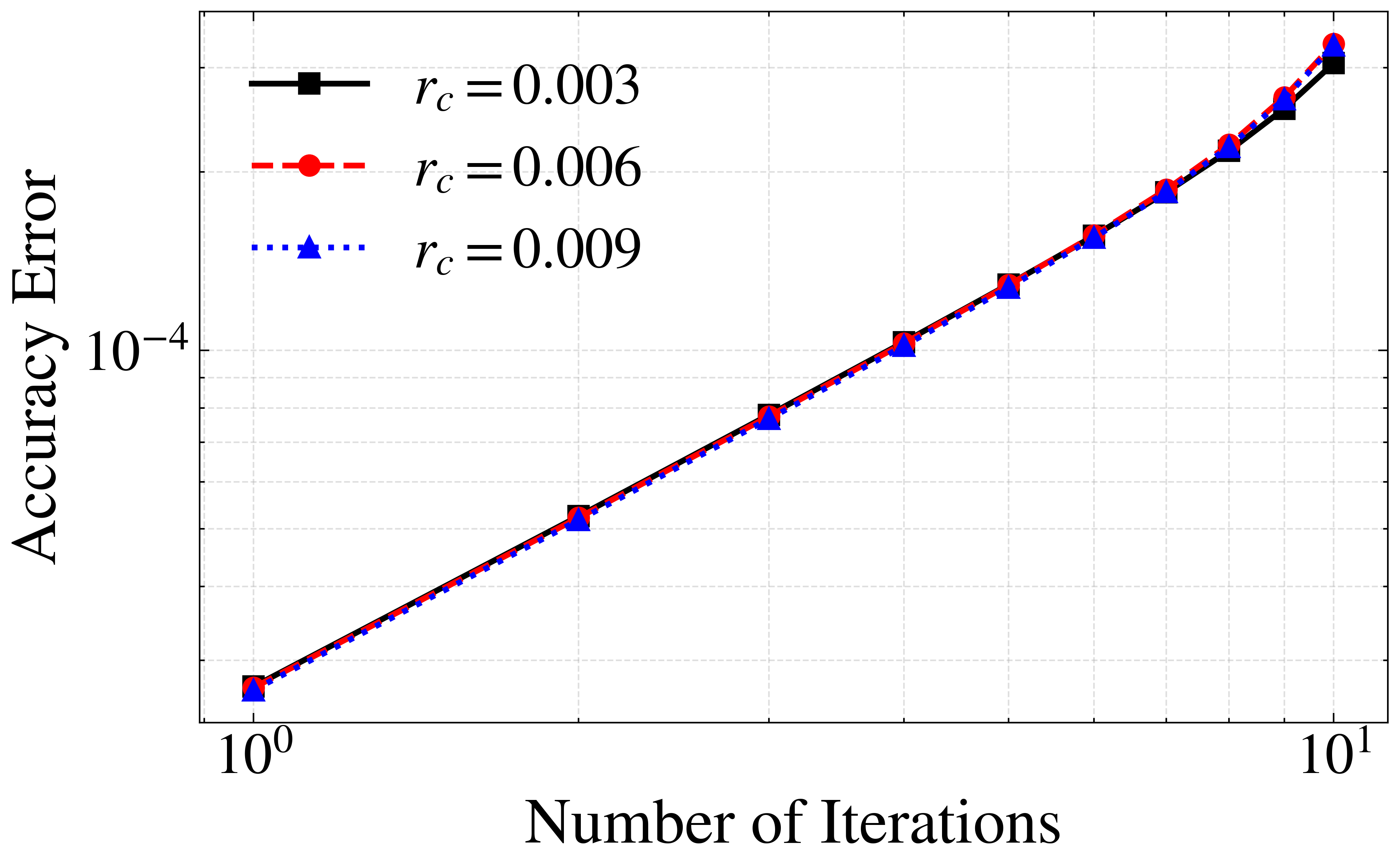}
    \caption{Accuracy}
    \end{subfigure}
    ~
    \begin{subfigure}[t]{0.48\linewidth}
    \centering
    \includegraphics[width=\linewidth]{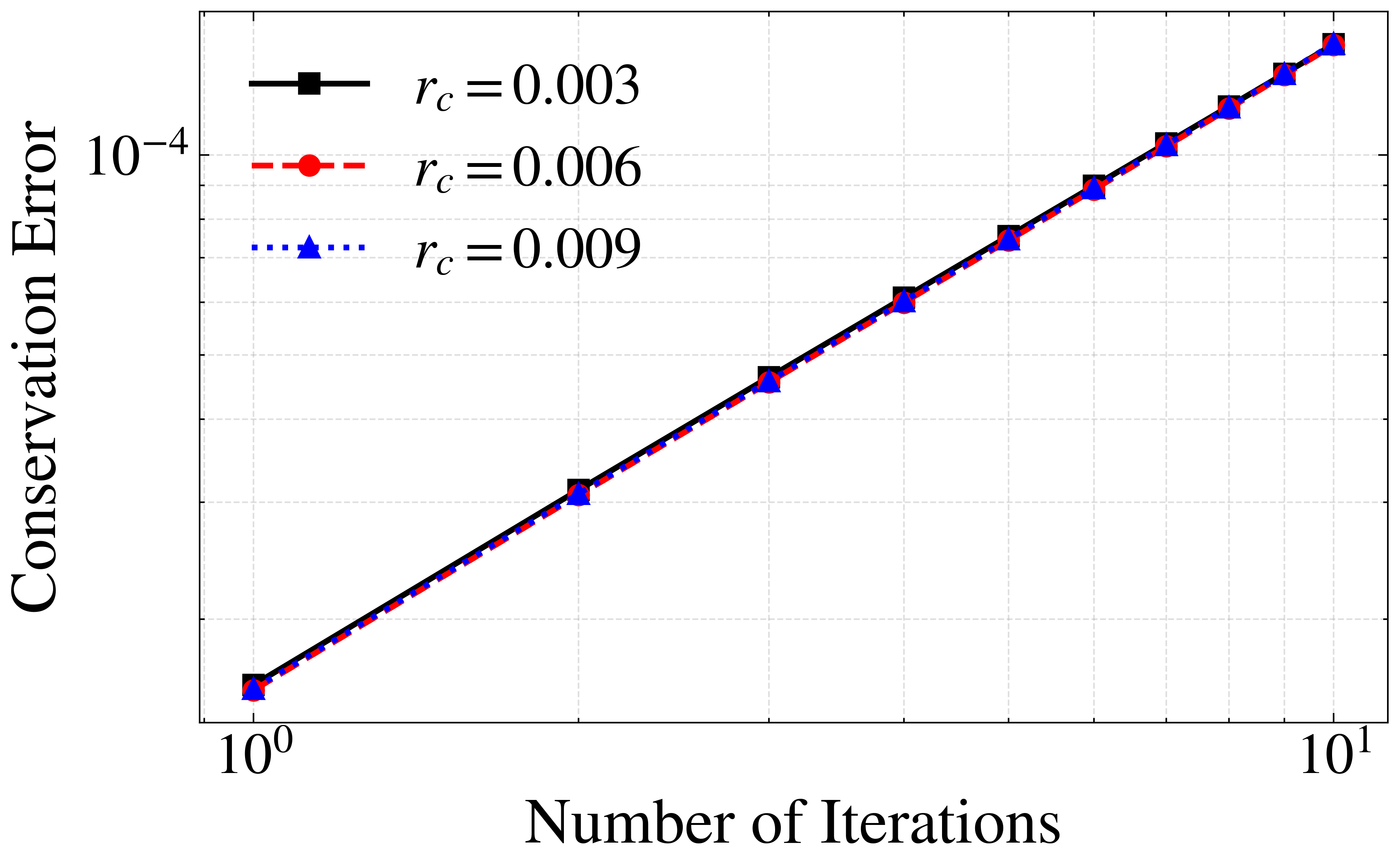}
    \caption{Conservation}
    \end{subfigure}
    \caption{Error of the quadratic field recovery when mapping field A from element centroids to vertices using C4 RBF.}
    \label{fig:quadratic-sin-cos}
\end{figure}

The accuracy error and conservation error are plotted as a function of iterations that map from the vertices of the linear triangle elements to the element centroids and back in an iterative process. One iteration corresponds to a complete cycle from vertices to centroid back to vertices such that the errors can be evaluated in a point-wise basis. This process is done on a production XGC mesh generated by TOMMS \cite{riaz_automated_2024} of the WEST tokamak reactor with 611,359 elements and 306,992 vertices (\Fig~\ref{fig:lcpp-meshes}a). The errors are dependent on the model geometry and field to be reconstructed.

To obtain an initial understanding of the errors associated with RBF field reconstruction, errors are computed with a range of cutoff radii with a smoothly varying function defined as
\begin{equation}
    f(x,y) = \sin(x)\cos(y)+2
    \label{eq:sin-cos-function}
\end{equation}
and shown in \Fig~\ref{fig:lcpp-meshes}b.

The resulting error for field mapping with the local polynomial fitting method described in section~\ref{sec:local-weighted-polynomial} with C4 radial basis functions are shown with a linear field recovery in \Fig~\ref{fig:linear-sincos} and with a quadratic field recovery in \Fig~\ref{fig:quadratic-sin-cos}. The recovered errors are similar for all cutoff radii tested for both the quadratic and linear cases.

One interesting observation is that although the quadratic reconstruction has initially better accuracy, the slope of the increase in error increases with the number of iterations. This is likely due to the poor conditioning of the Vandermonde matrix causing small errors in the reconstruction to amplify. Real coupling examples are often more forgiving than these iteration test cases because many solvers can damp out small errors in the coupled fields. However, each coupled set of applications is unique often requiring extensive experimentation to find coupling parameters that lead to a stable coupled simulation.

\subsubsection{Mesh Intersection Methods}\label{sec:mesh-intersection}
When mesh topology information is available, mesh intersection methods can be used. Mesh intersection methods are considered the gold standard for field mapping \cite{slatteryMeshfreeDataTransfer2016,jiaoCommonrefinementbasedDataTransfer2004} as they can utilize accurate \(L_2\) projection techniques over portions of the domain described by continuous functions. They are however complex to implement as they require the construction of element-element intersections. The key details of the technique are included here for completeness.

We seek to minimize the \(L_2\) error in the field transfer from \(f_s\) to \(f_t\) by solving:
\begin{equation}
    \pdv{}{f_{tA}} \int_{\Omega} (f_s-f_t)^2 \dd{\Omega} = \pdv{}{f_{tA}} \int_{\Omega} (f_s^2-2f_sf_t+f_t^2) \dd{\Omega}=0.
\end{equation}
Substituting in the definition of the target field \[f_t=\sum_{B=1}^n f_{tB} N_B \] we get
\begin{equation}
    \pdv{}{f_{tA}} \int_{\Omega} \left((\sum_{B=1}^n f_{tB} N_B)^2-2(\sum_{B=1}^n f_{tB} N_B)f_s+f_s^2\right) \dd{\Omega} = 0.
\end{equation}
And after differentiation
\begin{equation}
    \sum_{B=1}^n\int_{\Omega} N_A N_B\dd{\Omega}f_{tB} = \int_{\Omega} N_A f_s \dd{\Omega}.
\end{equation}
We immediately identify the integral \(M_{AB} = \int_{\Omega} N_A N_B\dd{\Omega}\) as the consistent mass matrix. We label the right hand side \(b_A\). This gives the sparse system to solve:
\begin{equation}
    \vb{M} \cdot \vb{f} = \vb{b}.
\end{equation}
that can be solved with any standard linear solver.

This is equivalent to the Galerkin projection
\begin{equation}
    \int_\Omega w f \dd{\Omega} = \int_\Omega w \phi \dd{\Omega}.
\end{equation}

The conservation error and accuracy will depend on how accurately the right-hand side integral, which is the product of a target field shape function and source field, can be computed. In the general case, the discretization of the source and target meshes are different, so the target shape functions may be discontinuous over a source element and likewise the source field may be discontinuous over the target element. This makes selecting a valid integration scheme challenging as using numerical quadrature requires smoothness that is violated if integrals are computed over either source or target meshes \cite{jiaoCommonrefinementbasedDataTransfer2004}. Instead, the source and target meshes are intersected to create a ``supermesh'' \cite{jiaoCommonrefinementbasedDataTransfer2004,farrellConservativeInterpolationUnstructured2009}. After intersection, the supermesh contains a set of polyhedra over which both source fields and target shape functions are smooth.

In PCMS, the mesh intersections are performed by a port of R3D \cite{powellR3dSoftwareFast2015} that has been adapted to work on the GPU and is included in Omega\_h\cite{ibanez2016conformal}. To compute integrals, a subdivision simplex mesh is constructed over each polyhedra. In the case where both the source and destination fields use Lagrange shape functions, the quadrature rules are selected such that they can exactly integrate a polynomial of the degree of the product of the two shape function definitions. In other cases where automatic selection is not possible, the user can provide an appropriate integration rule that maintains their desired level of accuracy.

\begin{figure}
\centering
 \begin{subfigure}[t]{0.48\linewidth}
     \centering
     \includegraphics[width=\linewidth]{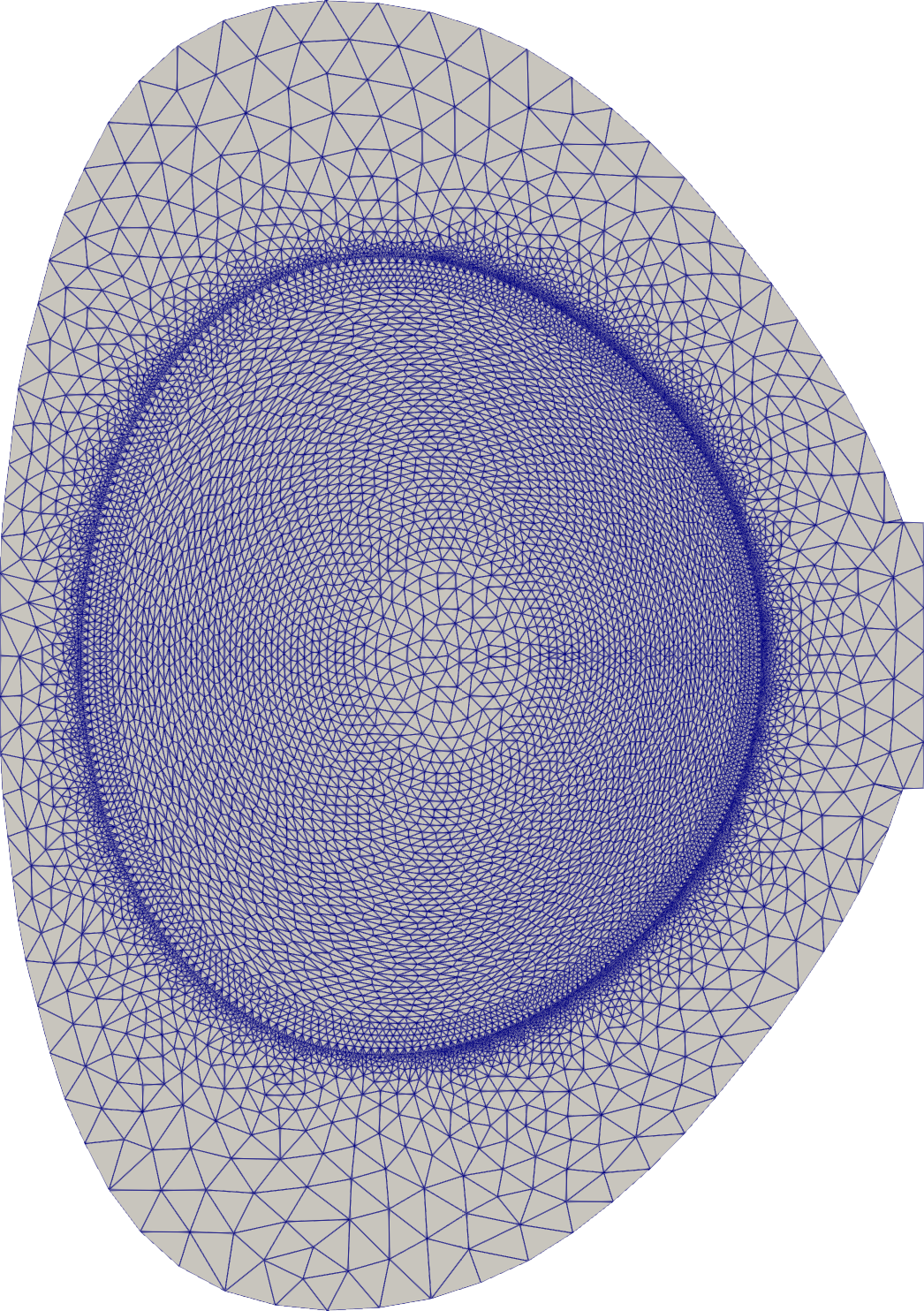}
     \caption{Mesh of LTX reactor used by the XGC code for evolving the plasma profiles.}
 \end{subfigure}
 ~
 \begin{subfigure}[t]{0.48\linewidth}
     \centering
     \includegraphics[width=\linewidth]{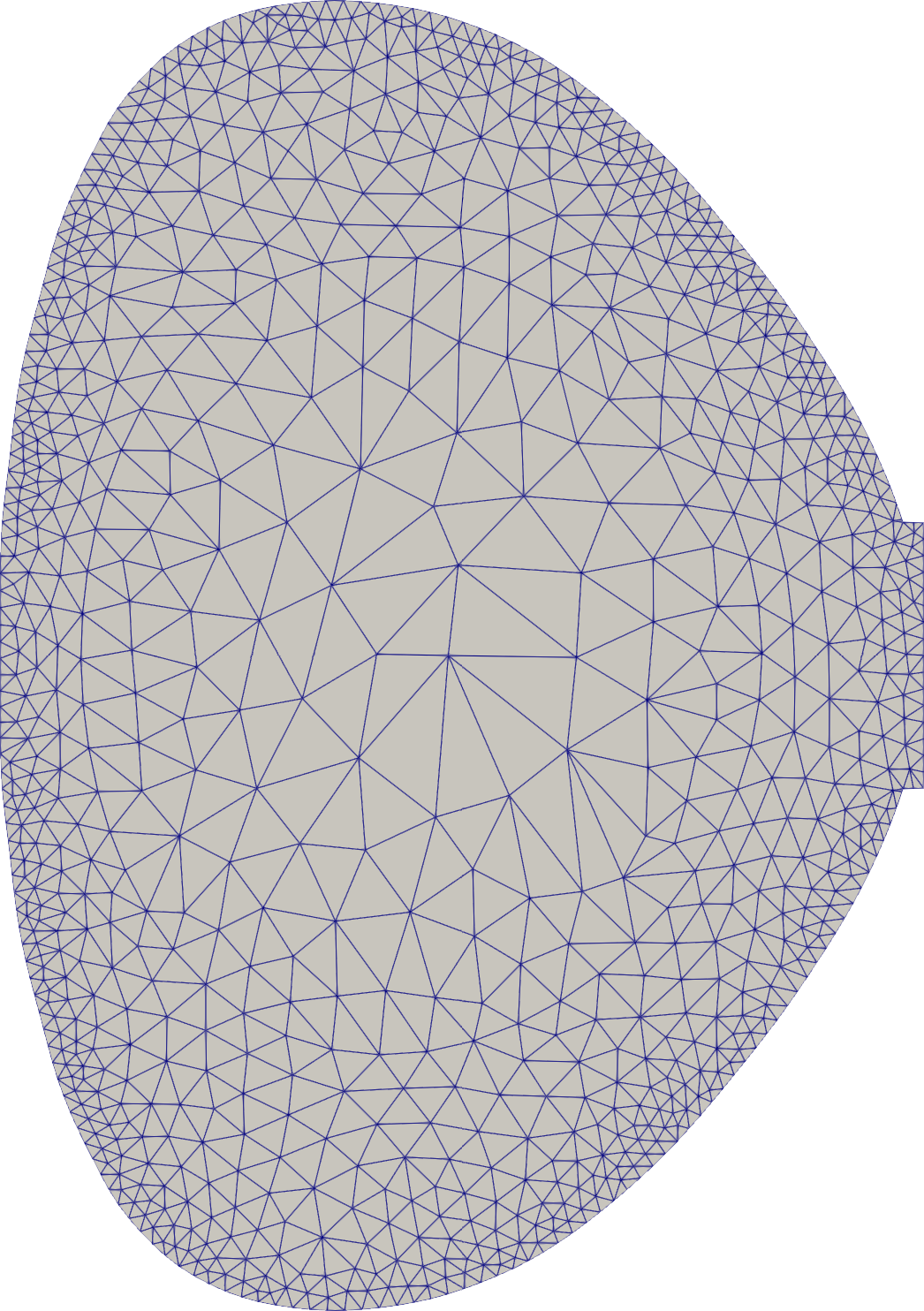}
     \caption{Mesh of the LTX reactor used by DEGAS2 to evolve the neutral particle profiles.}
 \end{subfigure}
 ~
 \begin{subfigure}[t]{0.48\linewidth}
     \centering
     \includegraphics[width=\linewidth]{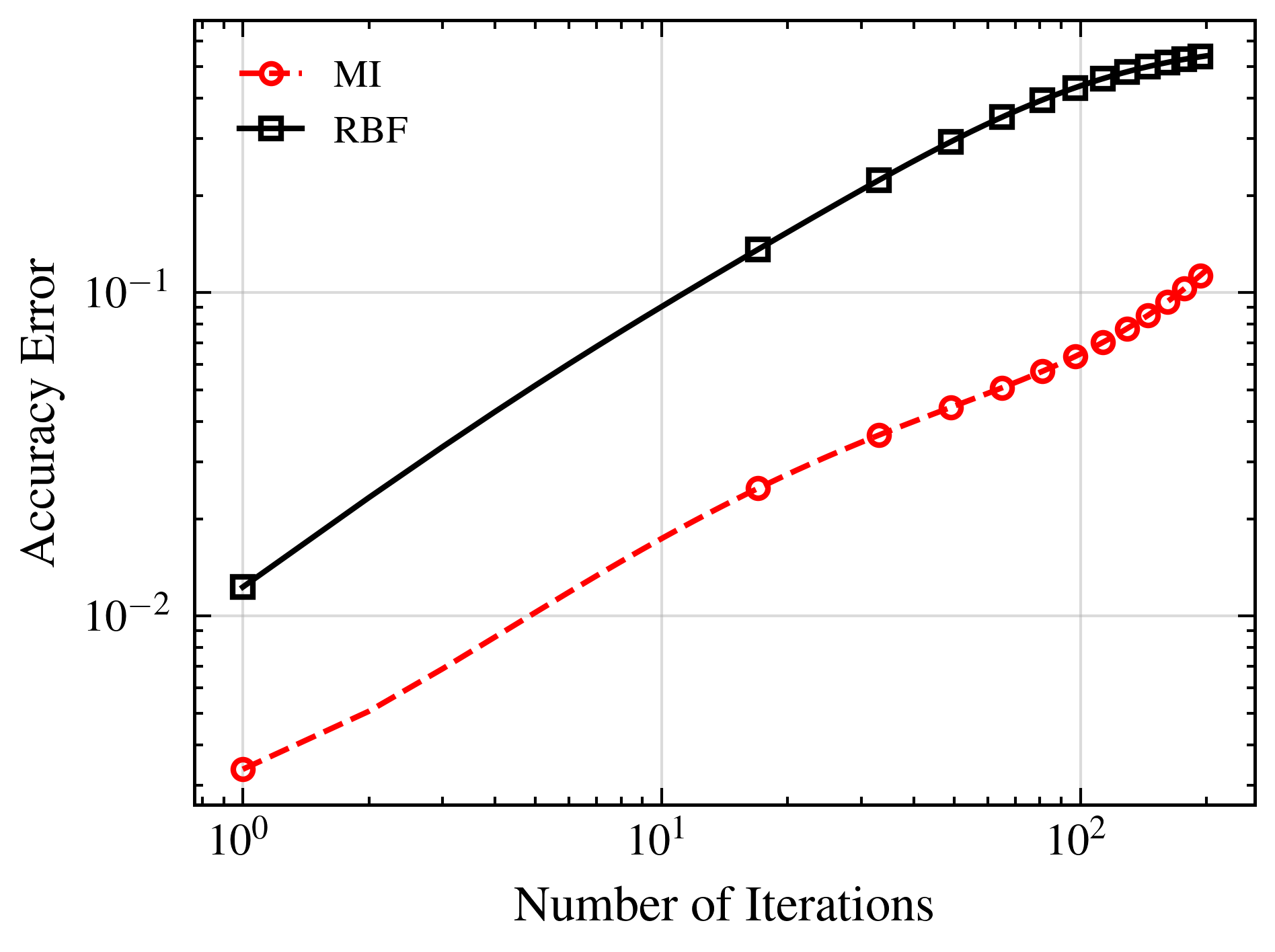}
     \caption{Accuracy of the field transfer methods.}
 \end{subfigure}
 ~
 \begin{subfigure}[t]{0.48\linewidth}
     \centering
     \includegraphics[width=\linewidth]{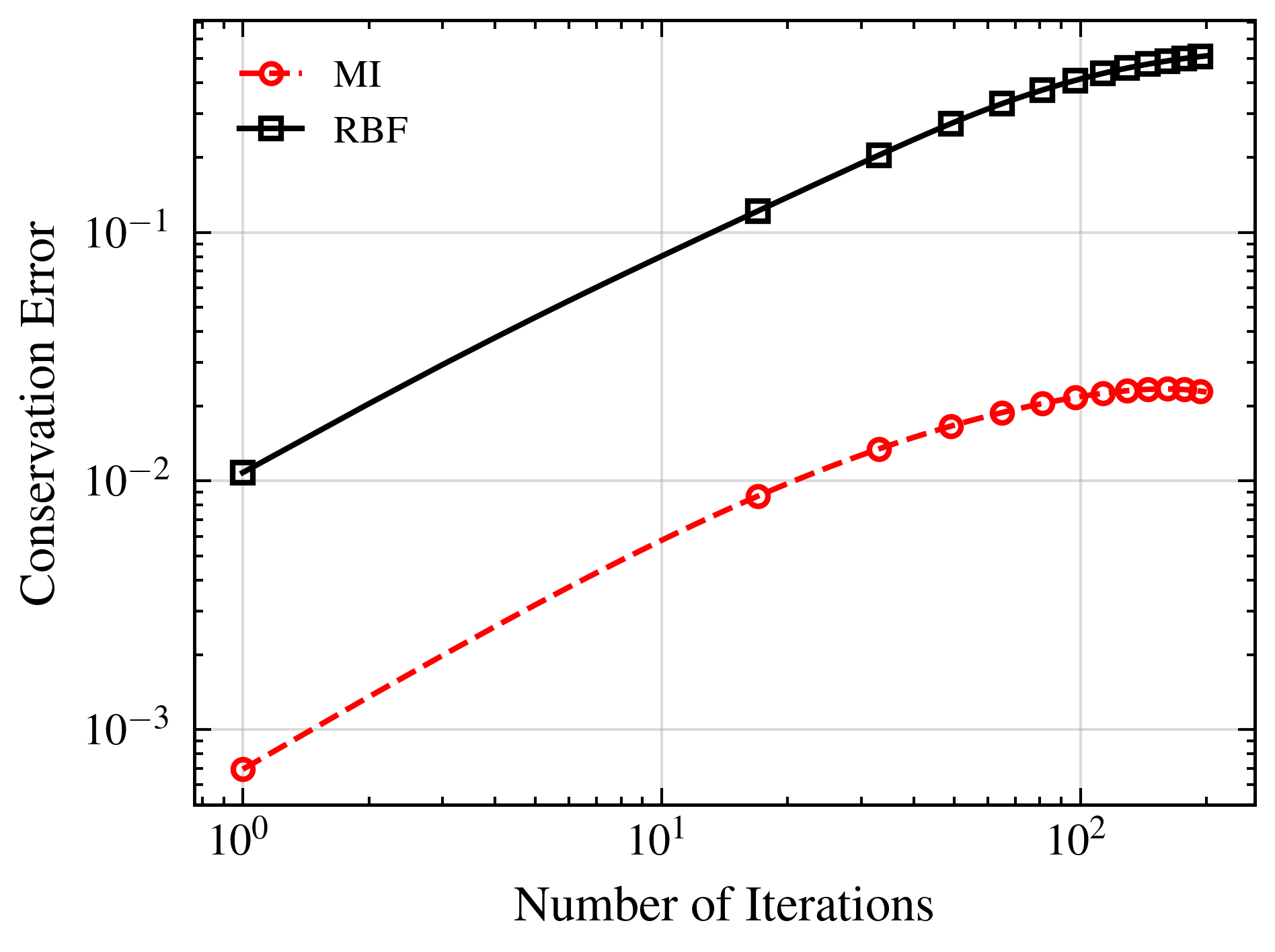}
     \caption{Conservation of the field transfer methods.}
 \end{subfigure}
 
 \caption{Evaluation of the  mesh intersection (MI) and C4 radial basis function (RBF) field transfer methods on the LTX reactor for a coupled gryokinetic and neutral particle recycling simulation.}
 \label{fig:xgc-degas2-conservative-coupling}
\end{figure}

The effectiveness of the mesh intersection methods are demonstrated in \Fig~\ref{fig:xgc-degas2-conservative-coupling}a-d which shows the accuracy (\Fig~\ref{fig:xgc-degas2-conservative-coupling}c)  and conservation error (\Fig~\ref{fig:xgc-degas2-conservative-coupling}d) obtained doing field transfer between a field-following mesh for the XGC gyrokinetic plasma code (\Fig~\ref{fig:xgc-degas2-conservative-coupling}a) and an unstructured mesh that is used for the DEGAS2 neutral particle transport code (\Fig~\ref{fig:xgc-degas2-conservative-coupling}b). In this example, the mesh intersection method is nearly an order of magnitude more accurate and more conservative than RBF, but the growth in the error versus field transfer iteration is similar.

\subsection{Point Localization}
Point localization is a fundamental component of all of the field transfer methods used by PCMS. It answers the following question: given a point in real coordinates which mesh element contains the point, and what are its parametric coordinate within the element. The naive approach to this problem scales as \(O(n*m)\) where \(n\) is the number of points to check and \(m\) is the number of elements within which it could reside. Typically, the discretization is held constant and a large number of points are queried which leads to the use of search structures that accelerate the query process with some up-front cost. There are a number of tree-based search structures such as quad and oct-trees \cite{jackinsOcttreesTheirUse1980}, KD-trees \cite{popovStacklessKDTreeTraversal2007}, R-trees \cite{beckmannRtreeEfficientRobust1990}, bounding volume hierarchies (BVH) \cite{lebrun-grandieArborXPerformancePortable2020,lauterbachFastBVHConstruction2009}, etc. Efforts to port these tree-based methods to GPUs have included the development of stackless KD-trees \cite{popovStacklessKDTreeTraversal2007,foleyKDtreeAccelerationStructures2005} and accelerated BVH searches that make use of purpose-built BVH GPU hardware for ray-tracing \cite{morricalAcceleratingUnstructuredMesh2022} or by making use of performance portability libraries \cite{lebrun-grandieArborXPerformancePortable2020}.

An alternative, and perhaps the simplest option is to use a uniform grid-based search structure. This option has seen wide adoption for a number of applications including predicates for polygon clipping \cite{menezesEmployingGPUsAccelerate2022}. Although the analysis is due for an update on modern hardware, \citeauthor{akmanGeometricComputingUniform1989} provides a reasonable basis that a single level uniform grid is sufficient unless there are orders of magnitude difference between the largest and smallest edge lengths \cite{akmanGeometricComputingUniform1989}.  
\citeauthor{zlatuskaRayTracingGpu2010} compared uniform grid to KD trees and BVH on a NVIDIA GTX 280 (released circa 2008) and found that the performance was workload dependent (in their case if they cast coherent or incoherant rays) \cite{zlatuskaRayTracingGpu2010}. \citeauthor{lubbeAnalysisParallelSpatial2020} investigated the use of uniform grids and BVH for GPU-based discrete element method (DEM) simulations finding that a Uniform Grid search uses less memory compared with a variant of the BVH and that each was preferred for a different part of their contact algorithm \cite{lubbeAnalysisParallelSpatial2020}.

A preliminary comparison of our GPU uniform-grid search with a state-of-the-art BVH library did not show a clear winner in terms of performance. Further investigation is critical, and we anticipate supporting both options in the future by making use of ArborX~\cite{lebrun-grandieArborXPerformancePortable2020} as a portable BVH backend.

For mesh-based operations, a strong understanding of element topology is often necessary for geometric consistency \cite{ibanez2016pumi}. For example, while performing 2D point-localization, a point may be located on an element face, edge, or vertex. If the localization library simply returns a predicate indicating that a point is \emph{in} (the closure of) an element, the answer can become inconsistent between simulations or MPI ranks when performed in parallel since there is no guaranteed order in which the elements are traversed. The solution is to make use of topology while performing localization so that the localization procedure indicates the lowest-order entity that the point could be classified on i.e., a point that is in the closure of an edge and a vertex will be classified on the vertex, not the edge. This requires any downstream algorithms that make use of localization information must be aware of how to work with the full topology information. This is the same strategy that has been used in the suite of SCOREC tools \cite{ibanez2016pumi,ibanez2016conformal,riaz2023modeling,mersonModeltraitsModelAttribute2021,diamond2021pumipic,hasanGPUAccelerationMonte2025}.

\subsection{Coordinate Transformations}
Commercial, mechanical and aerospace oriented, computer aided design (CAD) and computer-aided engineering (CAE) codes are based on two standardized coordinate systems, Cartesian and cylindrical. However, fusion codes employ a wide variety of coordinate systems that align to the physics of the problem \cite{dhaeseleerFluxCoordinatesMagnetic1991}. Two main approaches are taken; the first is to warp the coordinate frame such that the magnetic field lines are straight, but not necessarily aligned with one of the direction vectors. Examples include Boozer coordinates \cite{boozerPlasmaEquilibriumRational1981}, and PEST coordinates \cite{krugerRelationshipFluxCoordinates2019} applied in a range of codes \cite{linTurbulentTransportReduction1998,hirshmanSTEEPESTDESCENTMOMENT,mcmillanBEAMS3DNeutralBeam2014,jenkoElectronTemperatureGradient2000}. The second approach is to use a flux coordinate independent approach \cite{haririFluxcoordinateIndependentFieldaligned2013,stegmeirAdvancesFluxcoordinateIndependent2017} with structured meshes \cite{shanahanFluidSimulationsPlasma2018,michelsGENEXFullfGyrokinetic2021} and unstructured meshes \cite{kuNewHybridLagrangianNumerical2016, moritakaDevelopmentGyrokineticParticleCell2019} in a cylindrical or Cartesian coordinate system. In unstructured-mesh codes such as XGC, accuracy is maintained by placing degrees of freedom where the magnetic field lines intersect with poloidal planes \cite{kuNewHybridLagrangianNumerical2016, moritakaDevelopmentGyrokineticParticleCell2019,riaz2023modeling,Shephard_2024}.

Due to the anisotropy of the problem, many modern codes utilize a Fourier/spectral representations of the fields in some directions and spatial representations in the other directions \cite{sovinecNonlinearMagnetohydrodynamicsSimulation2004,candyHighaccuracyEulerianGyrokinetic2016}. In this setting, it is not desirable to first transform the toroidal coordinate to real space then use a Jacobian to transform to the target coordinate system because this requires the introduction of an intermediate spatial grid in the toroidal direction that can require significant memory, computation, and introduce errors. Instead, the coordinate transformation can often be written as a single operation from the source coordinate system to the destination coordinate system \cite{dominskiSpatialCouplingGyrokinetic2021,merloFirstCoupledGENE2021}. To support this, PCMS uses coordinate system strong-types that encode known transformations. This design is more flexible than an engineering-focused one which requires the definition of Jacobians for each set of coordinate systems to be transformed.

\section{Distributed Control/Communication} \label{sec:parallel-control}
Supporting the PCMS field transfer operations across multiple domains and decompositions requires a mechanism to efficiently coordinate the communication of unstructured data originating from unstructured meshes. The rendezvous algorithm coordinates the information passing between pairs of coupled parallel simulations ``when processors neither know which other processors to send data to, nor which other processors will be sending data to them''~\cite{plimptonRendezvousAlgorithmsLargescale2021}.
It is used in LAMMPS~\cite{plimptonRendezvousAlgorithmsLargescale2021}, the Parallel Unstructured Mesh Infrastructure (PUMI) for loading
multi-billion element meshes from file~\cite{rettenberger14}, and in the Data Transfer
Toolkit~\cite{slatteryDataTransferKit2013}.

This work extends the algorithm described in \cite{plimptonRendezvousAlgorithmsLargescale2021} to work with geometric model classification and other mesh partitioning schemes such as graph based partitioning and recursive coordinate bisection (RCB). The geometric model classification is the unique association of each mesh entity with the geometric model entity of equal or higher dimension that it lies on or within~\cite{beallShephardMeshDataStructure}.

PCMS also leverages the ADIOS2 library to enable communication within a single application or between multiple concurrently executing applications via a streaming data engine.
Use of an alternative DataSpaces \cite{docanDataSpacesInteractionCoordination2010} backend has been demonstrated as part of the Benesh integration \cite{davis2023benesh}. 
The communication layer that supports the rendezvous algorithm is general purpose and can be used as an independent library.

\begin{figure}
\centering
 \begin{subfigure}[t]{0.48\linewidth}
     \centering
     \includegraphics[width=\linewidth]{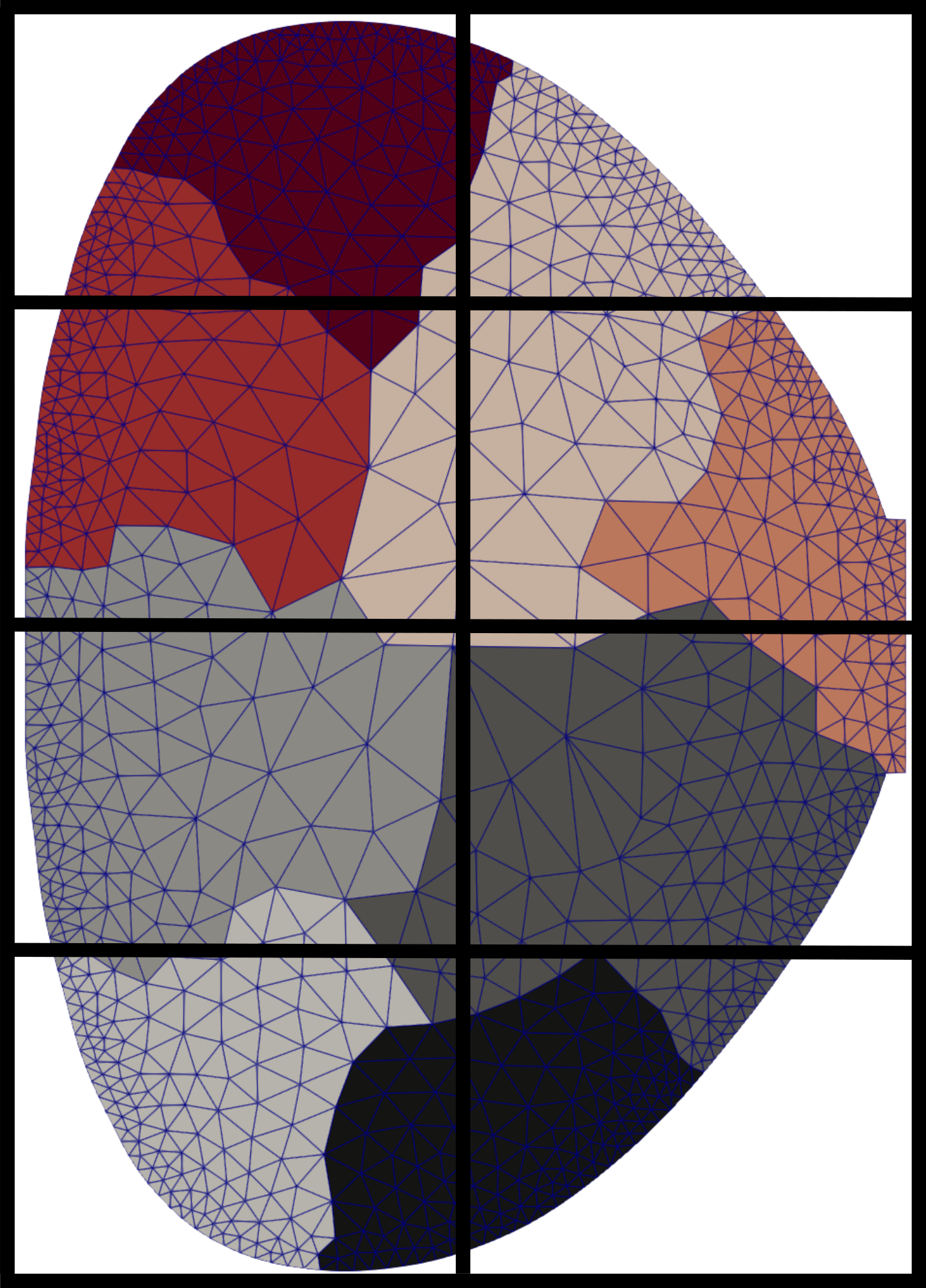}
     \caption{Unstructured mesh colored by its graph partition to eight processes.}
 \end{subfigure}
 ~
 \begin{subfigure}[t]{0.48\linewidth}
     \centering
     \includegraphics[width=\linewidth]{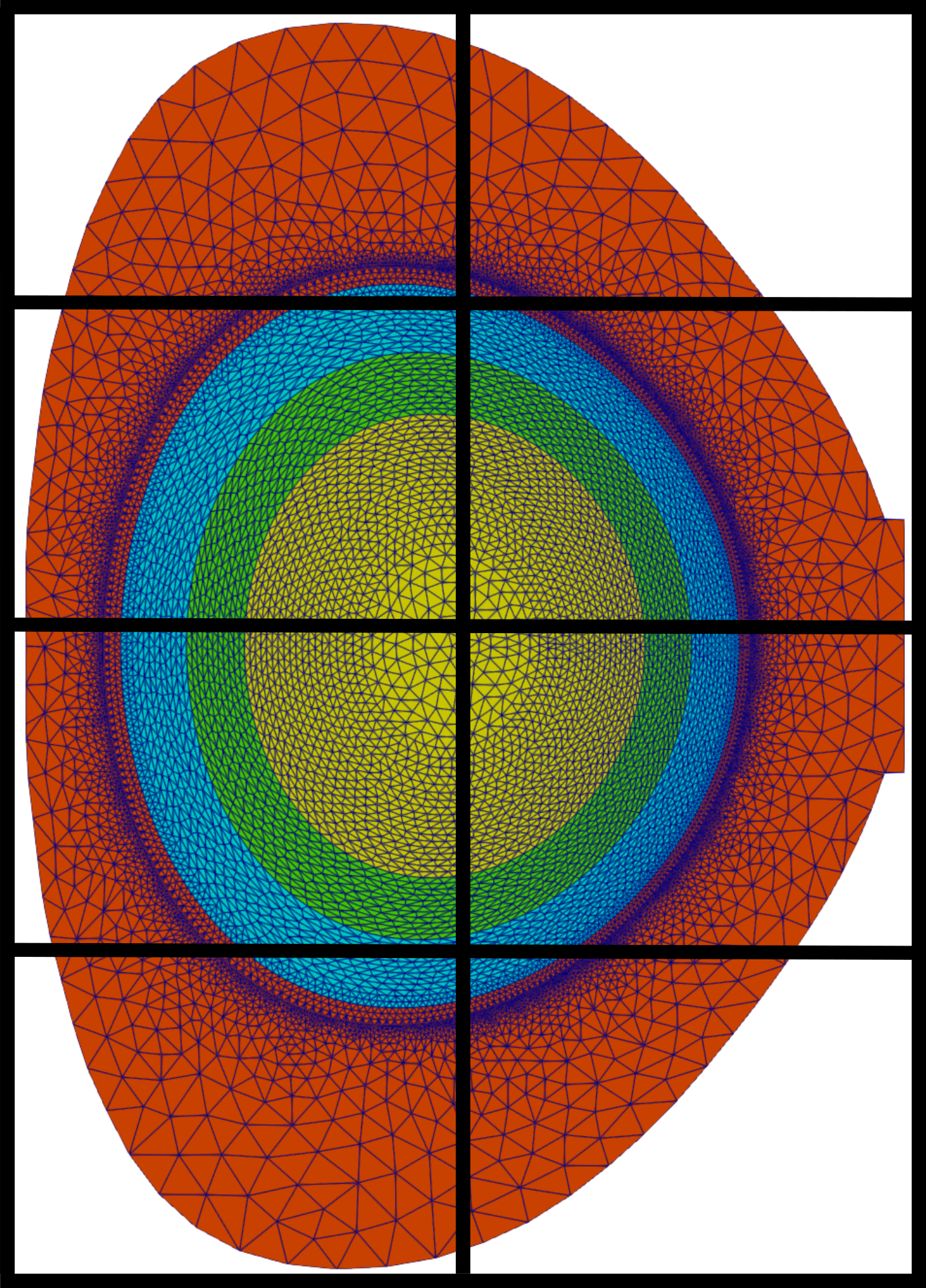}
     \caption{Field-following mesh colored by its classification-based partition to four processes.}
 \end{subfigure}
 ~
 \caption{Rendezvous partition (black grid) overlaid on unstructured and field-following LTX reactor meshes.}
 \label{fig:redevPartition}
\end{figure}

\Fig~\ref{fig:redevPartition} depicts examples of an unstructured and field-following mesh of the same domain
where information from one mesh is needed by the other.
Each mesh is colored by its native partition; a graph partition for the unstructured mesh and a geometric model classification-based partition for the field-following mesh.
Overlaid on both meshes is the black grid depicting the rendezvous partition used to
efficiently exchange data between the processes that do have the intersecting
portion of the domain.
The rendezvous partition allows processes that own data within a given mesh element to
`rendezvous' with the other processes that own data in that cell and perform
an exchange.

The key to the rendezvous algorithm is a partitioning of the common portion
of the domain that (1) has a relatively low memory usage and (2) supports computationally
efficient queries for membership within the partition given a point or mesh entity 
within the domain.
The structured grid from \Fig~\ref{fig:redevPartition} is one possible partition that
satisfies these requirements.

\section{Example Coupling Results} \label{sec:examples}
\subsection{Ion Density Field Mapping in Gyrokinetic Plasmas}
\begin{figure}
    \centering
    \includegraphics[width=0.5\linewidth]{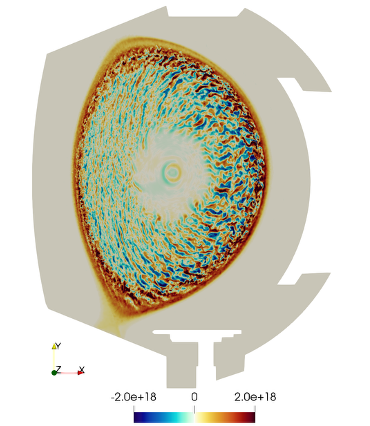}
    \caption{Ion density field from an XGCm simulation of the WEST tokamak.}
    \label{fig:ion-density}
\end{figure}
PCMS has been used to support a number of one-way and tightly coupled fusion simulations in up to five dimensions. This section provides a high level overview of some of these recent applications and the PCMS components that they make use of.

\Fig~\ref{fig:ion-density} shows a representative ion density field from an adiabatic electron \(\delta\)-f simulation of the plasma in the WEST reactor that was performed in the XGCm gyrokinetic microturbulence code \cite{zhang2023development}. XGCm employs the \(\delta\)-f gyrokinetic particle-in-cell methodology which is a hybrid particle and mesh simulation technique. In XGCm, weighted marker particles evolve the deviation of the five-dimensional distribution function from an analytic Maxwellian distribution. The particle data is integrated in velocity space and represented on 2D poloidal plane meshes at fixed toroidal angles using the elementary toroidal coordinate system in a 3D configuration space.

The specifics of the WEST analysis case are omitted as the goal of this example is not to convey any physical insight, but to provide an example test case for the PCMS field transfer operators that operate on production data.

\begin{figure}
    \centering
    \begin{subfigure}[t]{0.48\linewidth}
    \centering
    \includegraphics[width=\linewidth]{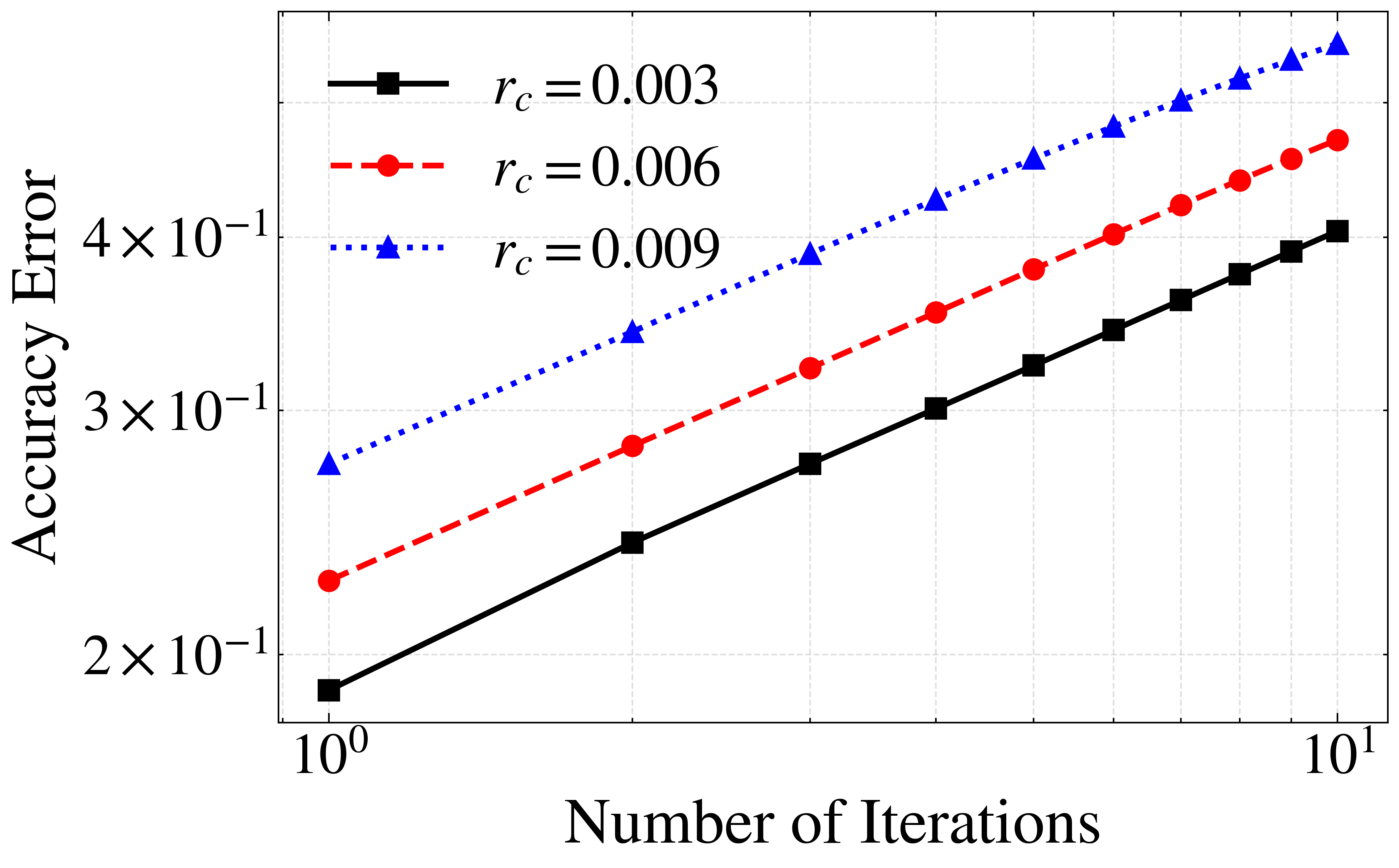}
    \caption{Accuracy}
    \end{subfigure}
    ~
    \begin{subfigure}[t]{0.48\linewidth}
    \centering
    \includegraphics[width=\linewidth]{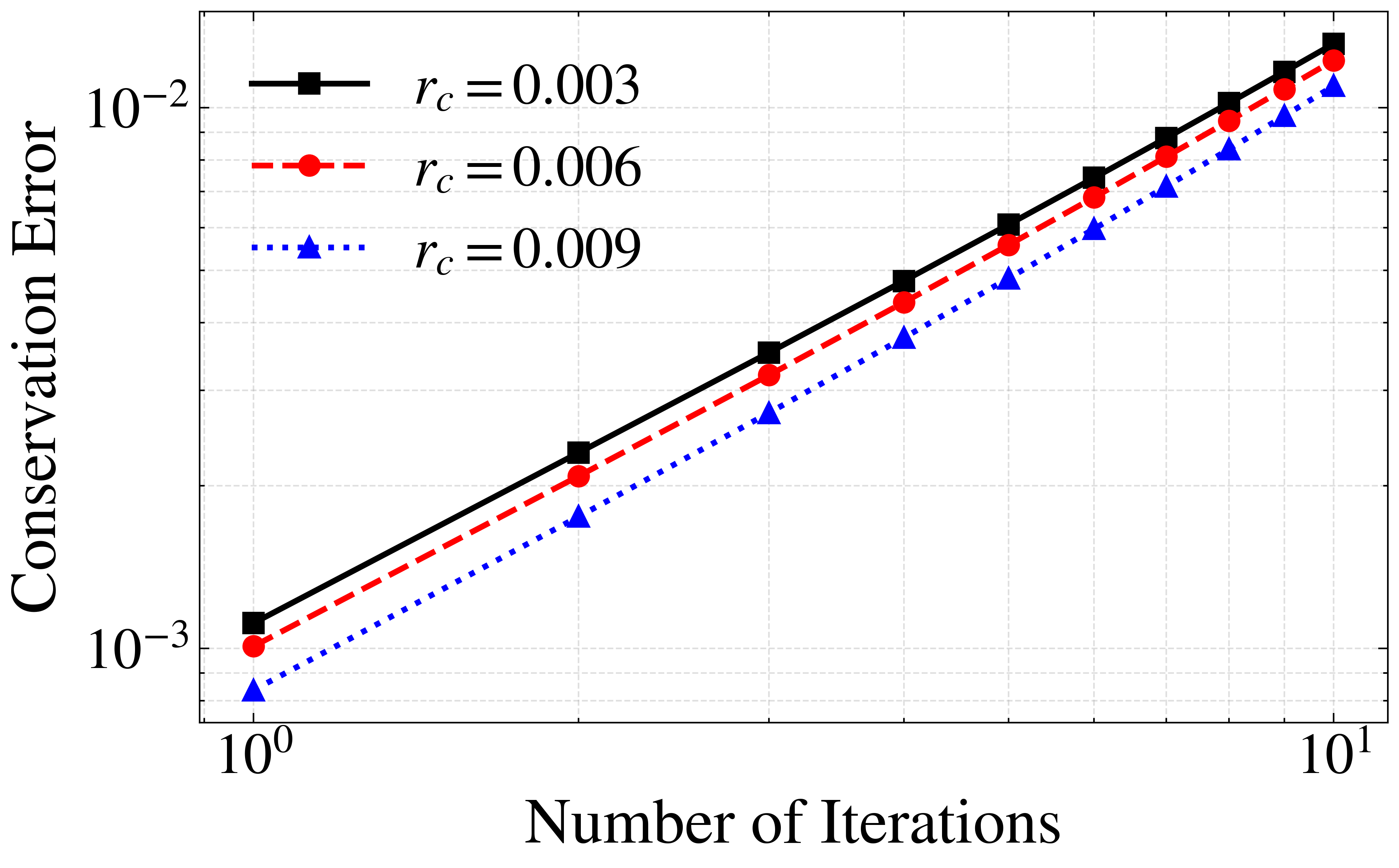}
    \caption{Conservation}
    \end{subfigure}
    \caption{Error of the linear field recovery when mapping the ion density field from element centroids to vertices using C4 RBF.}
    \label{fig:linear-ion-density}
\end{figure}

\begin{figure}
    \centering
    \begin{subfigure}[t]{0.48\linewidth}
    \centering
    \includegraphics[width=\linewidth]{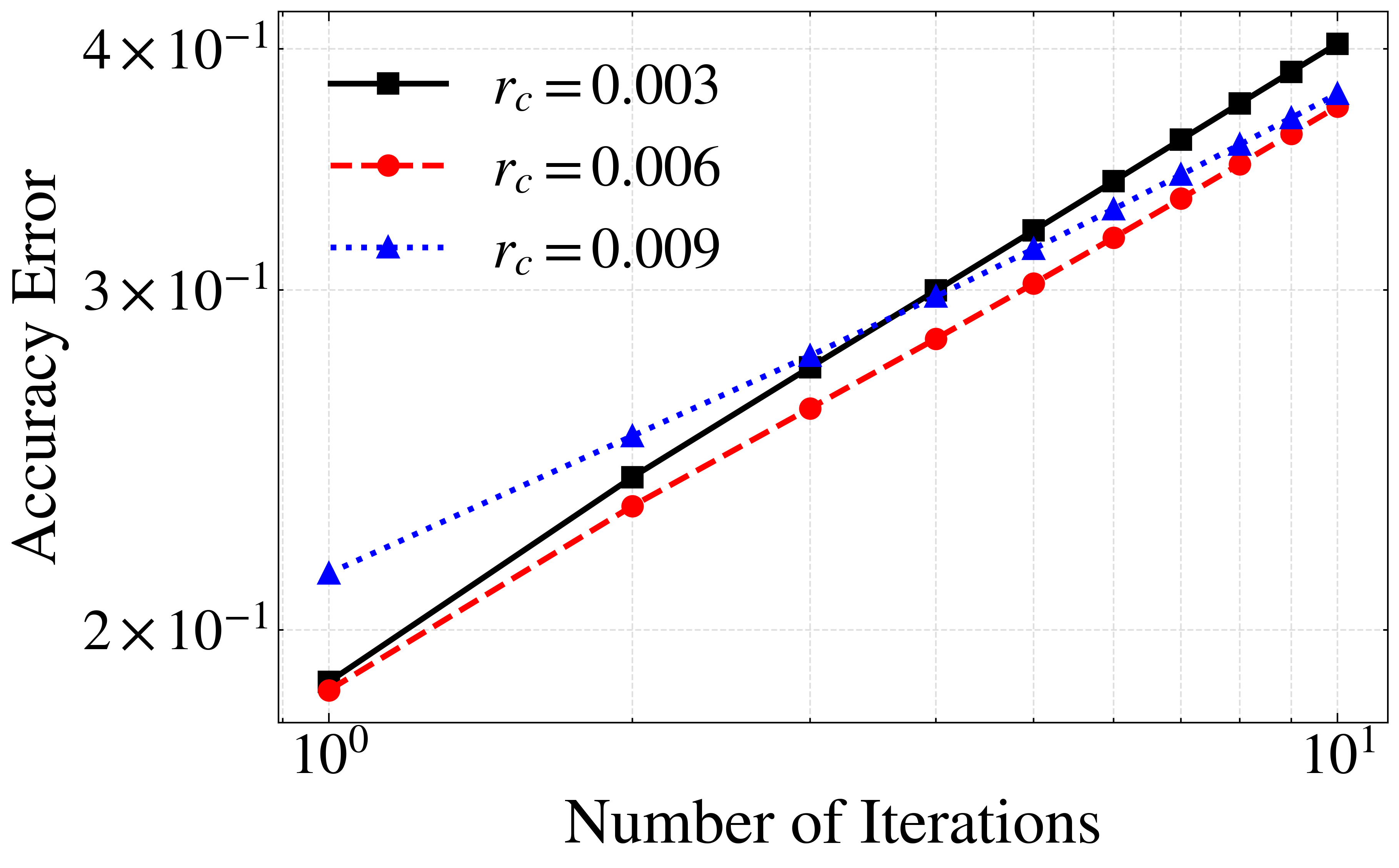}
    \caption{Accuracy}
    \end{subfigure}
    ~
    \begin{subfigure}[t]{0.48\linewidth}
    \centering
    \includegraphics[width=\linewidth]{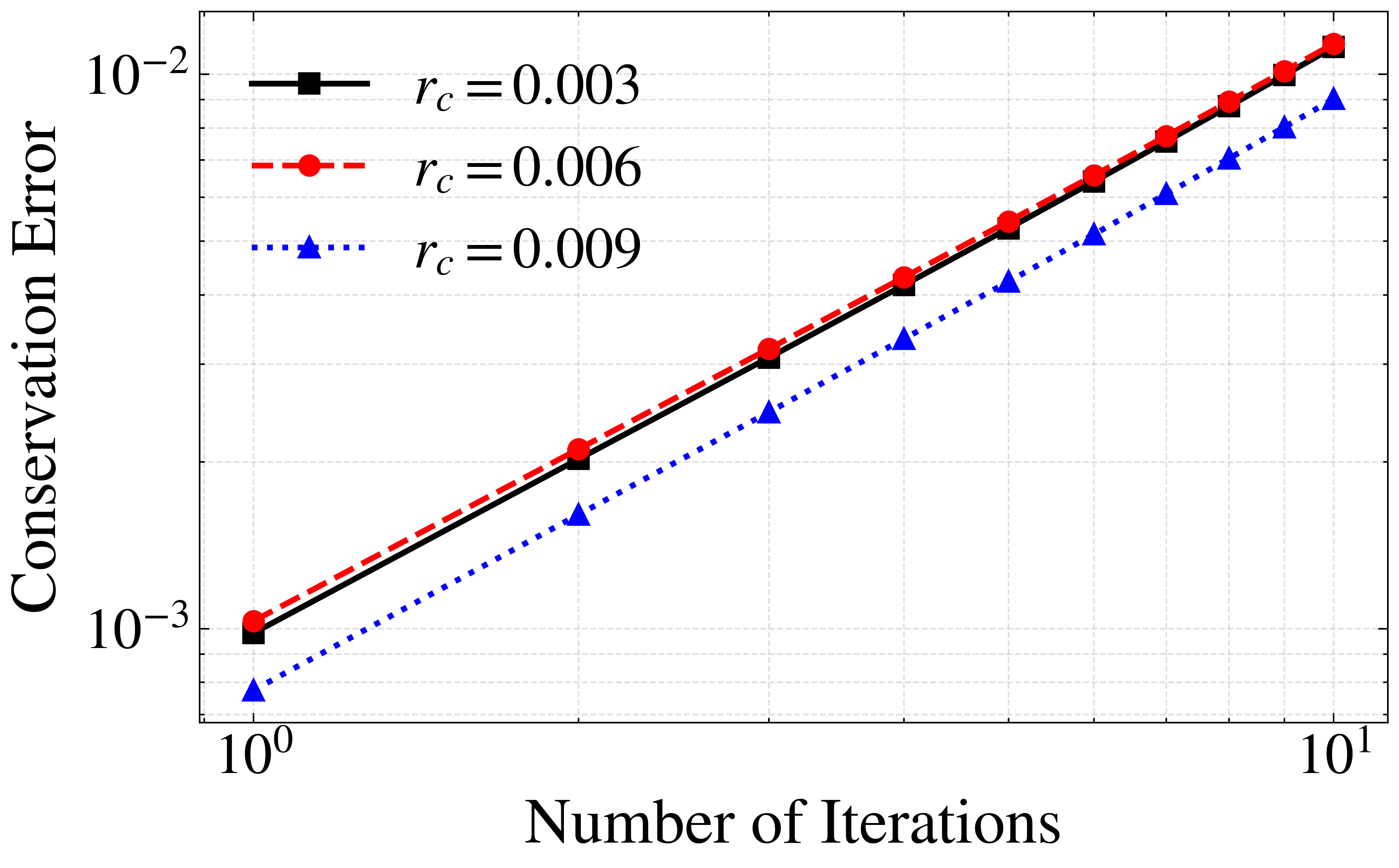}
    \caption{Conservation}
    \end{subfigure}
    \caption{Error of the quadratic field recovery when mapping the ion density field from element centroids to vertices using C4 RBF.}
    \label{fig:quadratic-ion-density}
\end{figure}

\Figs~\ref{fig:linear-ion-density} and \ref{fig:quadratic-ion-density} give the accuracy error and conservation error that is obtained by mapping the ion density field (\Fig~\ref{fig:ion-density}) from the vertices of the linear triangle elements to the element centroids and back in an iterative process that is identical to that used to generate \Fig~\ref{fig:linear-sincos} and \Fig~\ref{fig:quadratic-sin-cos}.

In \Fig~\ref{fig:linear-ion-density} the linear polynomial basis is used to recover the field values. Both the accuracy error and conservation error increase with the number of iterations. The error is observably different for each value of the cutoff radius. This is different than the linear accuracy error for the analytic function (equation~\ref{eq:sin-cos-function}) shown in \Fig~\ref{fig:linear-sincos}. The accuracy errors reported for the ion field are three orders of magnitude worse than for the the analytic field and the conservation errors are about an order of magnitude worse. We believe this is because the ion density field has much finer oscillating structures rather than global trends.

\Fig~\ref{fig:quadratic-ion-density} shows the errors associated with the ion field recovery with a quadratic polynomial basis. The errors for each choice of cutoff radius are consistent and similar to the best-choice of the cutoff radius for the linear case.

\subsection{5D Coupling of GNET and GTC}
Energetic particle transport is important for optimizing Stellarator and Tokamak reactor performance and is performed with specialized codes such as GNET \cite{murakami5DSimulationStudy2000} or BEAMS3D \cite{mcmillanBEAMS3DNeutralBeam2014} that simulate the drift kinetic equations in 5D phase space. The resulting 5D slowing down distributions are a desired input condition to gyrokinetic plasma simulations in codes such as the Gyrokinetic Toroidal Code (GTC) that solves the gyrokinetic \(\delta\)-f or total-f equations \cite{linTurbulentTransportReduction1998}. 

\begin{figure}
\centering
 \begin{subfigure}[t]{0.48\linewidth}
     \centering
     \includegraphics[width=\linewidth]{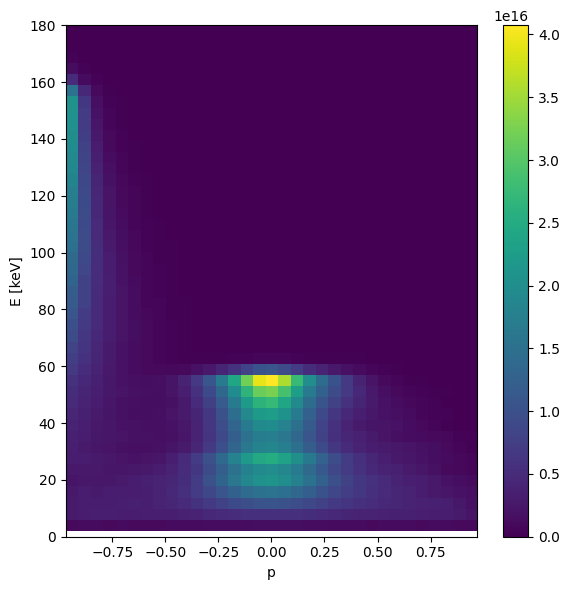}
     \caption{Original GNET distribution function.}
 \end{subfigure}
 ~
 \begin{subfigure}[t]{0.48\linewidth}
     \centering
     \includegraphics[width=\linewidth]{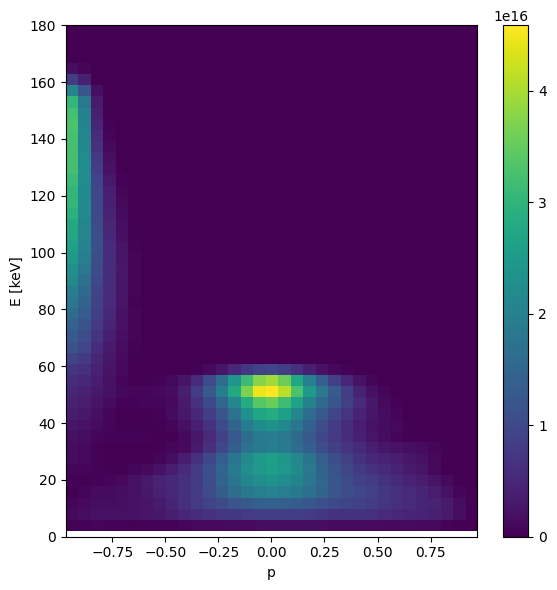}
     \caption{Sampled GTC distribution function using 15` million sample points.}
 \end{subfigure}
 \caption{Comparison of the velocity space histogram of the 5D particle distribution function used for coupling GNET neutral beam injection to GTC microturbulence simulations.}
 \label{fig:5d-distribution-function}
\end{figure}

PCMS was used to reconstruct the 5D GNET distribution function and was sampled to construct a valid input distribution function for GTC. \Fig~\ref{fig:5d-distribution-function} provides a 2D histogram of the distribution function in velocity space after integrating over the three phase-space dimensions. \Fig~\ref{fig:5d-distribution-function}a is the original distribution function from GNET and \Fig~\ref{fig:5d-distribution-function}b is the distribution function of 15 million marker particles that have been sampled from a recovered 5D representation of the GNET distribution function.

It is worth noting that the recovered particle distribution function is slightly more diffuse than the original distribution function, however it largely retains the desired shape. When used as inputs to GTC, the recovered distribution function results in a stable solve and relatively smooth lost-particle functions.

\section{Performance and Scaling} \label{sec:performance}
\subsection{Performance results}
Measuring the performance of tightly coupled simulations in PCMS presents a unique challenge because each application and the coupler run as independent applications without a shared synchronization method. On large supercomputers, each node's clock is not guaranteed to be synchronized with a level of accuracy acceptable for computing timing measurements. To overcome these challenges, timing results are measured in the coupling server over ten rounds, where each communication round represents sending data to the applications, and receiving data back from the applications. In other words, timing results are measured from the initiation of the first send in the coupler to the last receive in the coupler.

\begin{figure}
    \centering
    \includegraphics[width=\linewidth]{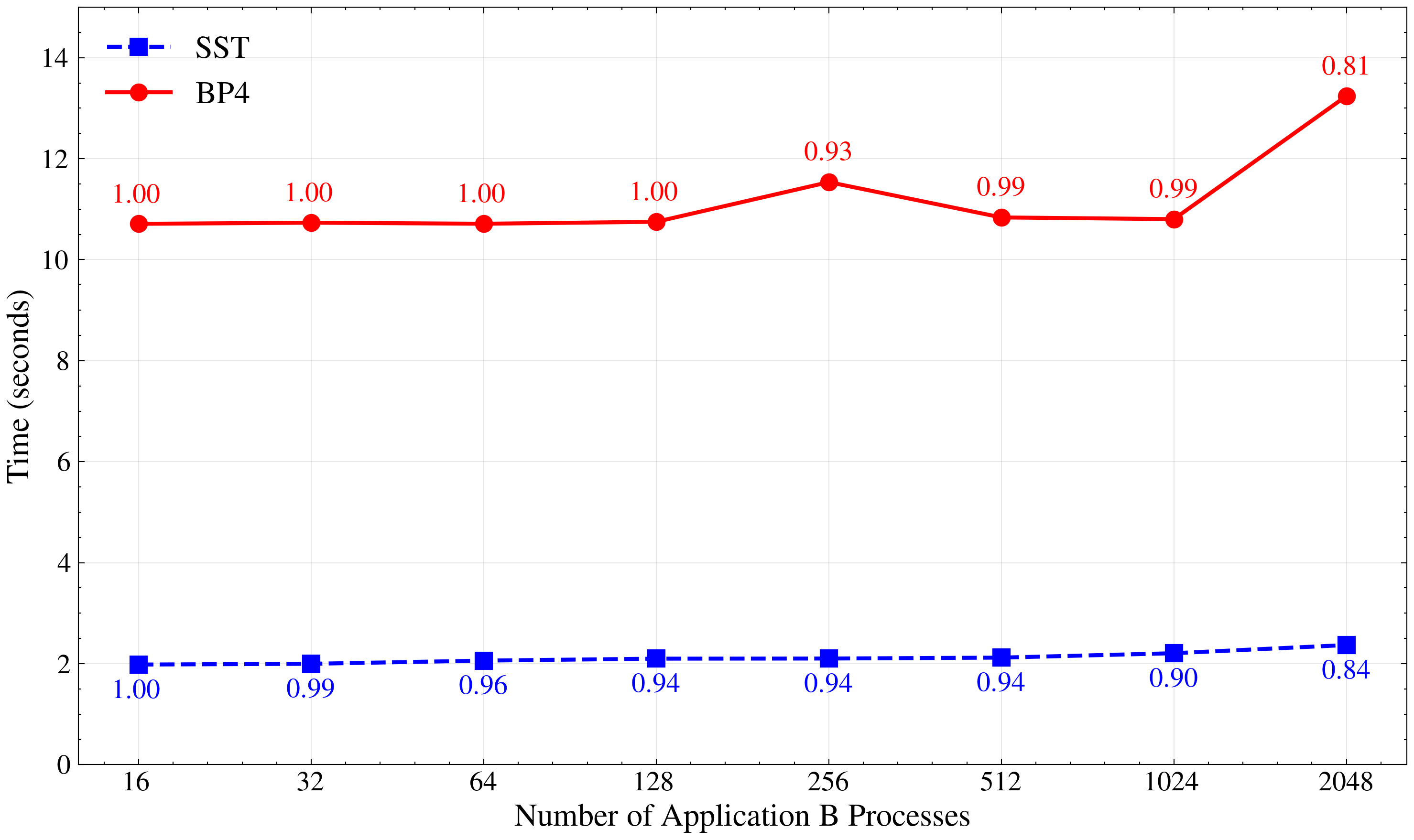}
    \caption{PCMS weak scaling study on full overlap proxy test running on up to 260 nodes (256 nodes for application B and two nodes each for the coupler and application A) on OLCF Frontier using the ADIOS2 BP4 and SST IO Engines.  
    The data points in each series are labeled with their weak scaling efficiency versus the baseline for that Engine at 16 processes; a value of one indicates 100\% efficiency and higher is better.}
    \label{fig:frontier-weak-scaling}
\end{figure}

Weak scaling tests are performed with two applications and a coupling server. Application A and the coupler both utilize 16 processes for the entire scaling study. The number of processes in application B are scaled up to 2048 processes. The weak scaling results on Frontier, a HPE Cray EX supercomputer located at the Oak Ridge Leadership Computing Facility, are shown in \Fig~\ref{fig:frontier-weak-scaling} using the ADIOS2 BP4 and SST IO Engines\footnote{timing data and job launch sripts available here: \url{https://doi.org/10.5281/zenodo.16989048}}. Each Frontier node has four AMD250X each with two Graphics Compute Dies (GCDs), one 64-core AMD EPYC 7763 64 core CPU and 512 GB of DDR4 memory. We use one process per GCD (i.e., application A and the coupler both run on two nodes and application B runs on up to 256 nodes).

For weak scaling, the coupler utilizes a geometric classification based partitioning scheme similar to that shown in \Fig~\ref{fig:redevPartition}b on a mesh of the D3D reactor with four million triangle elements. This geometric classification-based partition is convenient because for core-edge integration, the overlap region is defined by a set of geometric model faces and their closure. Both applications utilize a geometric, recursive inertial bisection~\cite{williamsRIB}, based partitioning scheme.

Weak scaling results for the ADIOS2 BP4 file-based engine are shown in \Fig~\ref{fig:frontier-weak-scaling}. The scaling results are fairly flat until application B hits 1024 processes, at which point there is a significant increase in the runtime. This is likely due to limitations of the parallel file system.

The ADIOS2 SST engine utilizes remote direct memory access (RDMA) through libfabric.
The coupling runtime with the SST engine is 5.6 times shorter than with BP4 at the largest process count (2048 application B processes) and is at least 4.9 times shorter at lower counts.
As shown in \Fig~\ref{fig:frontier-weak-scaling} via data point labels, the weak scaling efficiency was also better with SST as the time increase from 16 to 2048 processes was 19\% versus BP4's increase of 23\% (13.24s/10.7s).

\section{Integration with other Tools} \label{sec:integration}
PCMS has been integrated with a number of other software packages to provide additional functionality and enhance user experience. The current integrations fit into two categories: workflow management, and physics based interfaces that provide a more natural point-of-entry for domain experts.

Currently, PCMS works with two workflow management layers: EFFIS \cite{suchyta2022exascale} which is agnostic to the application specifics, but handles job scheduling on leadership class supercomputers and provides some utilities for in-situ monitoring and analysis. PCMS has also been integrated with Benesh \cite{davis2023benesh} which utilizes code annotations to support asynchronous coordination of field information.

The interfaces in PCMS are designed to be general enough to support a wide range of applications. However this level of abstraction introduces its own set of jargon and code conventions that add complexity for domain specialists. Additionally, specific use-cases may benefit from additional assumptions or relaxed performance constraints that can lead to further simplified interfaces. To support the fusion community, our team has been coordinating with the ADIOS2 team to interact with the IMAS/OMAS data schema \cite{meneghiniNeuralnetworkAcceleratedCoupled2021}. PCMS has also been integrated into FusionIO which provides readers for a number of fusion-relevant formats such as M3D-C\textsuperscript{1}, mars, geqdsk, and gato.

\section{Conclusion and Future Work} \label{sec:future-work}
This paper introduces the Parallel Coupler for Multimodel Simulations (PCMS). PCMS aims to fill a critical gap in generalized coupling tools needed for complex physics applications. It targets volume coupling for high-dimensional problems such as kinetic simulations in 6D and gyrokinetic microturbulence simulations in 5D and can work with complex coordinate systems and geometries defined by physics or engineering CAD packages.

PCMS supports a variety of field transfer algorithms, ranging from those that can operate without knowledge of source or target discretizations or fields to those that utilize full discretization and field information to maintain physical constraints. These field transfer operations are intended to be fast and are performed on GPUs (including the mesh intersections and point localizations).

In addition to field transfer methods, PCMS supports distributed control and communication methods. These methods afford the ability to couple unstructured field information from applications that have arbitrary runtime-defined partitioning schemes. The effectiveness of this strategy was demonstrated on Frontier.

Work in progress in PCMS includes adding automatic time step control through an integration with SUNDIALS \cite{hindmarsh2005sundials,gardner2022sundials}, lifting operators to support coupling of axisymmetric 2D fluid and 5D gyrokinetic codes, and additional fusion relevant field definitions such as those used in M3D-C\textsuperscript{1} and NIMROD.

\section{Declaration of Competing Interest}
The authors declare that they have no known competing financial interests or personal relationships that could have appeared to influence the work reported in this paper.

\section{Acknowledgment}
This research was supported by the U.S. Department of Energy, Office of
Science, under awards DE-SC0021285 (FASTMath SciDAC Institute) and
DE-AC02-09CH11466 (StellFoundry: High-fidelity Digital Models for
Fusion Pilot Plant Design). This work used the resources of 
the Oak Ridge Leadership Computing Facility at the Oak Ridge National Laboratory, which is supported by the Office of Science of the U.S. Department of Energy under Contract No. DE-AC05-00OR22725, and 
the National Energy Scientific Computing Center (NERSC) at the Lawrence Berkeley National Laboratory (awards FES-ERCAP-m4505, FES-ERCAP-m4564).
Any opinions, findings, and conclusions or recommendations expressed in this material are those of the author(s) 
and do not necessarily reflect the views of the U.S. Department of Energy. We gratefully acknowledge use of the research computing resources of the Empire AI Consortium, Inc, with support from the State of New York, the Simons Foundation, and the Secunda Family Foundation.

\bibliography{unstr-mesh-fusion-bibliography}

\begin{thebibliography}{86}
\providecommand{\natexlab}[1]{#1}
\providecommand{\url}[1]{\texttt{#1}}
\expandafter\ifx\csname urlstyle\endcsname\relax
  \providecommand{\doi}[1]{doi: #1}\else
  \providecommand{\doi}{doi: \begingroup \urlstyle{rm}\Url}\fi

\bibitem[Bungartz et~al.(2015)Bungartz, Lindner, Mehl, and
  Uekermann]{bungartzPlugandplayCouplingApproach2015}
Hans-Joachim Bungartz, Florian Lindner, Miriam Mehl, and Benjamin Uekermann.
\newblock A plug-and-play coupling approach for parallel multi-field
  simulations.
\newblock \emph{Computational Mechanics}, 55:\penalty0 1119--1129, 2015.

\bibitem[Chourdakis et~al.(2022)Chourdakis, Davis, Rodenberg, Schulte, Simonis,
  Uekermann, Abrams, Bungartz, Cheung~Yau, Desai, Eder, Hertrich, Lindner,
  Rusch, Sashko, Schneider, Totounferoush, Volland, Vollmer, and
  Koseomur]{chourdakisPreCICEV2Sustainable2022}
Gerasimos Chourdakis, Kyle Davis, Benjamin Rodenberg, Miriam Schulte,
  Fr\'ed\'eric Simonis, Benjamin Uekermann, Georg Abrams, Hans-Joachim
  Bungartz, Lucia Cheung~Yau, Ishaan Desai, Konrad Eder, Richard Hertrich,
  Florian Lindner, Alexander Rusch, Dmytro Sashko, David Schneider, Amin
  Totounferoush, Dominik Volland, Peter Vollmer, and Oguz~Ziya Koseomur.
\newblock {{preCICE}} v2: {{A}} sustainable and user-friendly coupling library.
\newblock \emph{Open Research Europe}, 2:\penalty0 51, 2022.

\bibitem[Gaston et~al.(2009)Gaston, Newman, Hansen, and
  Lebrun-Grandie]{gastonMOOSEParallelComputational2009}
Derek Gaston, Chris Newman, Glen Hansen, and Damien Lebrun-Grandie.
\newblock {{MOOSE}}: A parallel computational framework for coupled systems of
  nonlinear equations.
\newblock \emph{Nuclear Engineering and Design}, 239\penalty0 (10):\penalty0
  1768--1778, 2009.

\bibitem[Slattery et~al.(2013)Slattery, Wilson, and
  Pawlowski]{slatteryDataTransferKit2013}
S~R Slattery, P~P~H Wilson, and R~P Pawlowski.
\newblock The data transfer kit: A geometric rendezvous-based tool for
  multiphysics data transfer.
\newblock In \emph{International Conference on Mathematics and Computational
  Methods Applied to Nuclear Science \& Engineering (M\&C 2013)}, page~11, Sun
  Valley, ID, May 2013. American Nuclear Society.

\bibitem[ORN(2025)]{ORNLCEESDataTransferKit2025}
{ORNL}-{CEES}/{DataTransferKit}, July 2025.
\newblock URL \url{https://github.com/ORNL-CEES/DataTransferKit}.
\newblock original-date: 2014-10-15T17:35:07Z.

\bibitem[noa(2025)]{noauthor_laristraportage_2025}
laristra/portage, March 2025.
\newblock URL \url{https://github.com/laristra/portage}.
\newblock original-date: 2016-12-21T18:17:59Z.

\bibitem[Shephard et~al.(2024)Shephard, Merson, Sahni, Castillo, Joshi, Nath,
  Riaz, Seol, Smith, Zhang, Beall, Klaas, Nastasia, and
  Tendulkar]{Shephard_2024}
Mark~S. Shephard, Jacob Merson, Onkar Sahni, Angel~E. Castillo, Aditya~Y.
  Joshi, Dhyanjyoti~D. Nath, Usman Riaz, E.~Seegyoung Seol, Cameron~W. Smith,
  Chonglin Zhang, Mark~W. Beall, Ottmar Klaas, Rocco Nastasia, and Saurabh
  Tendulkar.
\newblock Unstructured mesh tools for magnetically confined fusion system
  simulations.
\newblock \emph{Engineering with Computers}, 40\penalty0 (5):\penalty0
  3319--3336, April 2024.
\newblock ISSN 1435-5663.
\newblock \doi{10.1007/s00366-024-01976-2}.
\newblock URL \url{http://dx.doi.org/10.1007/s00366-024-01976-2}.

\bibitem[Fish et~al.(2021)Fish, Wagner, and
  Keten]{fishMesoscopicMultiscaleModelling2021}
Jacob Fish, Gregory~J. Wagner, and Sinan Keten.
\newblock Mesoscopic and multiscale modelling in materials.
\newblock \emph{Nature Materials}, 20\penalty0 (6):\penalty0 774--786, June
  2021.
\newblock ISSN 1476-4660.
\newblock \doi{10.1038/s41563-020-00913-0}.

\bibitem[Tadmor and Miller(2011)]{tadmorModelingMaterialsContinuum2011}
Ellad~B. Tadmor and Ronald~E. Miller.
\newblock \emph{Modeling Materials: Continuum, Atomistic, and Multiscale
  Techniques}.
\newblock Cambridge University Press, Cambridge ; New York, 2011.
\newblock ISBN 978-0-521-85698-0.

\bibitem[Keyes et~al.(2013)Keyes, McInnes, Woodward, Gropp, Myra, Pernice,
  Bell, Brown, Clo, Connors, Constantinescu, Estep, Evans, Farhat, Hakim,
  Hammond, Hansen, Hill, Isaac, Jiao, Jordan, Kaushik, Kaxiras, Koniges, Lee,
  Lott, Lu, Magerlein, Maxwell, McCourt, Mehl, Pawlowski, Randles, Reynolds,
  Rivi{\`e}re, R{\"u}de, Scheibe, Shadid, Sheehan, Shephard, Siegel, Smith,
  Tang, Wilson, and Wohlmuth]{keyesMultiphysicsSimulationsChallenges2013}
David~E Keyes, Lois~C McInnes, Carol Woodward, William Gropp, Eric Myra,
  Michael Pernice, John Bell, Jed Brown, Alain Clo, Jeffrey Connors, Emil
  Constantinescu, Don Estep, Kate Evans, Charbel Farhat, Ammar Hakim, Glenn
  Hammond, Glen Hansen, Judith Hill, Tobin Isaac, Xiangmin Jiao, Kirk Jordan,
  Dinesh Kaushik, Efthimios Kaxiras, Alice Koniges, Kihwan Lee, Aaron Lott,
  Qiming Lu, John Magerlein, Reed Maxwell, Michael McCourt, Miriam Mehl, Roger
  Pawlowski, Amanda~P Randles, Daniel Reynolds, Beatrice Rivi{\`e}re, Ulrich
  R{\"u}de, Tim Scheibe, John Shadid, Brendan Sheehan, Mark Shephard, Andrew
  Siegel, Barry Smith, Xianzhu Tang, Cian Wilson, and Barbara Wohlmuth.
\newblock Multiphysics simulations: Challenges and opportunities.
\newblock \emph{The International Journal of High Performance Computing
  Applications}, 27\penalty0 (1):\penalty0 4--83, February 2013.
\newblock ISSN 1094-3420, 1741-2846.
\newblock \doi{10.1177/1094342012468181}.

\bibitem[Tadmor et~al.(1996)Tadmor, Ortiz, and
  Phillips]{tadmorQuasicontinuumAnalysisDefects1996}
E.~B. Tadmor, M.~Ortiz, and R.~Phillips.
\newblock Quasicontinuum analysis of defects in solids.
\newblock \emph{Philosophical Magazine A}, 73\penalty0 (6):\penalty0
  1529--1563, June 1996.
\newblock ISSN 0141-8610.
\newblock \doi{10.1080/01418619608243000}.

\bibitem[Wagner and Liu(2003)]{wagnerCouplingAtomisticContinuum2003}
Gregory~J. Wagner and Wing~Kam Liu.
\newblock Coupling of atomistic and continuum simulations using a bridging
  scale decomposition.
\newblock \emph{Journal of Computational Physics}, 190\penalty0 (1):\penalty0
  249--274, September 2003.
\newblock ISSN 0021-9991.
\newblock \doi{10.1016/S0021-9991(03)00273-0}.

\bibitem[Fish et~al.(2007)Fish, Nuggehally, Shephard, Picu, Badia, Parks, and
  Gunzburger]{fishConcurrentAtCCoupling2007}
Jacob Fish, Mohan~A. Nuggehally, Mark~S. Shephard, Catalin~R. Picu, Santiago
  Badia, Michael~L. Parks, and Max Gunzburger.
\newblock Concurrent atc coupling based on a blend of the continuum stress and
  the atomistic force.
\newblock \emph{Computer Methods in Applied Mechanics and Engineering},
  196\penalty0 (45--48):\penalty0 4548--4560, September 2007.
\newblock ISSN 0045-7825.
\newblock \doi{10.1016/j.cma.2007.05.020}.

\bibitem[Xu and Gracie(2010)]{xuConcurrentCouplingAtomistic2010}
Mei Xu and R~Gracie.
\newblock Concurrent coupling of atomistic and continuum models.
\newblock In T~Belytschko, editor, \emph{Multiscale Methods: Bridging the
  Scales in Science and Eng.}, pages 93--133. oxford university press, 2010.

\bibitem[Miller and Tadmor(2009)]{millerUnifiedFrameworkPerformance2009}
Ronald~E Miller and E~B Tadmor.
\newblock A unified framework and performance benchmark of fourteen multiscale
  atomistic/continuum coupling methods.
\newblock \emph{Modelling and Simulation in Materials Science and Engineering},
  17\penalty0 (5):\penalty0 053001, July 2009.
\newblock ISSN 0965-0393, 1361-651X.
\newblock \doi{10.1088/0965-0393/17/5/053001}.

\bibitem[Parks et~al.(2008)Parks, Bochev, and
  Lehoucq]{parksConnectingAtomistictoContinuumCoupling2008}
Michael~L. Parks, Pavel~B. Bochev, and Richard~B. Lehoucq.
\newblock Connecting atomistic-to-continuum coupling and domain decomposition.
\newblock \emph{Multiscale Modeling \& Simulation}, 7\penalty0 (1):\penalty0
  362--380, January 2008.
\newblock ISSN 1540-3459.
\newblock \doi{10.1137/070682848}.

\bibitem[Yin et~al.(2022)Yin, Zhang, Yu, and
  Karniadakis]{yinInterfacingFiniteElements2022}
Minglang Yin, Enrui Zhang, Yue Yu, and George~Em Karniadakis.
\newblock Interfacing finite elements with deep neural operators for fast
  multiscale modeling of mechanics problems.
\newblock \emph{Computer Methods in Applied Mechanics and Engineering},
  402:\penalty0 115027, December 2022.
\newblock ISSN 0045-7825.
\newblock \doi{10.1016/j.cma.2022.115027}.

\bibitem[Dominski et~al.(2021)Dominski, Cheng, Merlo, Carey, Hager, Ricketson,
  Choi, Ethier, Germaschewski, Ku, Mollen, Podhorszki, Pugmire, Suchyta,
  Trivedi, Wang, Chang, Hittinger, Jenko, Klasky, Parker, and
  Bhattacharjee]{dominskiSpatialCouplingGyrokinetic2021}
J.~Dominski, J.~Cheng, G.~Merlo, V.~Carey, R.~Hager, L.~Ricketson, J.~Choi,
  S.~Ethier, K.~Germaschewski, S.~Ku, A.~Mollen, N.~Podhorszki, D.~Pugmire,
  E.~Suchyta, P.~Trivedi, R.~Wang, C.~S. Chang, J.~Hittinger, F.~Jenko,
  S.~Klasky, S.~E. Parker, and A.~Bhattacharjee.
\newblock Spatial coupling of gyrokinetic simulations, a generalized scheme
  based on first-principles.
\newblock \emph{Physics of Plasmas}, 28\penalty0 (2):\penalty0 022301, February
  2021.
\newblock ISSN 1070-664X, 1089-7674.
\newblock \doi{10.1063/5.0027160}.

\bibitem[Merlo et~al.(2021)Merlo, Janhunen, Jenko, Bhattacharjee, Chang, Cheng,
  Davis, Dominski, Germaschewski, Hager, Klasky, Parker, and
  Suchyta]{merloFirstCoupledGENE2021}
G.~Merlo, S.~Janhunen, F.~Jenko, A.~Bhattacharjee, C.~S. Chang, J.~Cheng,
  P.~Davis, J.~Dominski, K.~Germaschewski, R.~Hager, S.~Klasky, S.~Parker, and
  E.~Suchyta.
\newblock First coupled gene--xgc microturbulence simulations.
\newblock \emph{Physics of Plasmas}, 28\penalty0 (1):\penalty0 012303, January
  2021.
\newblock ISSN 1070-664X, 1089-7674.
\newblock \doi{10.1063/5.0026661}.

\bibitem[Novak et~al.(2024)Novak, Brooks, Shriwise, and
  Davis]{novakMonteCarloMultiphysics2024}
A.J. Novak, H.~Brooks, P.~Shriwise, and A.~Davis.
\newblock Monte carlo multiphysics simulation on adaptive unstructured mesh
  geometry.
\newblock \emph{Nuclear Engineering and Design}, 429:\penalty0 113589, December
  2024.
\newblock ISSN 00295493.
\newblock \doi{10.1016/j.nucengdes.2024.113589}.

\bibitem[Weiler(1985)]{weiler1985edge}
Kevin Weiler.
\newblock Edge-based data structures for solid modeling in curved-surface
  environments.
\newblock \emph{IEEE Computer graphics and applications}, 5\penalty0
  (1):\penalty0 21--40, 1985.

\bibitem[Beall et~al.(2004)Beall, Walsh, and Shephard]{beall2004comparison}
Mark~W Beall, Joe Walsh, and Mark~S Shephard.
\newblock A comparison of techniques for geometry access related to mesh
  generation.
\newblock \emph{Engineering with Computers}, 20\penalty0 (3):\penalty0
  210--221, 2004.

\bibitem[Godoy et~al.(2020)Godoy, Podhorszki, Wang, Atkins, Eisenhauer, Gu,
  Davis, Choi, Germaschewski, Huck, Huebl, Kim, Kress, Kurc, Liu, Logan, Mehta,
  Ostrouchov, Parashar, Poeschel, Pugmire, Suchyta, Takahashi, Thompson,
  Tsutsumi, Wan, Wolf, Wu, and Klasky]{godoyADIOSAdaptableInput2020}
William~F. Godoy, Norbert Podhorszki, Ruonan Wang, Chuck Atkins, Greg
  Eisenhauer, Junmin Gu, Philip Davis, Jong Choi, Kai Germaschewski, Kevin
  Huck, Axel Huebl, Mark Kim, James Kress, Tahsin Kurc, Qing Liu, Jeremy Logan,
  Kshitij Mehta, George Ostrouchov, Manish Parashar, Franz Poeschel, David
  Pugmire, Eric Suchyta, Keichi Takahashi, Nick Thompson, Seiji Tsutsumi,
  Lipeng Wan, Matthew Wolf, Kesheng Wu, and Scott Klasky.
\newblock Adios 2: The adaptable input output system. a framework for
  high-performance data management.
\newblock \emph{SoftwareX}, 12, July 2020.
\newblock ISSN 2352-7110.
\newblock \doi{10.1016/j.softx.2020.100561}.

\bibitem[Suchyta et~al.(2022)Suchyta, Klasky, Podhorszki, Wolf, Chang, Choi,
  Davis, Dominski, Ethier, Shephard, et~al.]{suchyta2022exascale}
Eric Suchyta, Scott Klasky, Norbert Podhorszki, Matthew Wolf, CS~Chang, Jong
  Choi, Philip~E Davis, Julien Dominski, St{\'e}phane Ethier, Mark~S. Shephard,
  et~al.
\newblock The exascale framework for high fidelity coupled simulations
  {(EFFIS)}: Enabling whole device modeling in fusion science.
\newblock \emph{The International Journal of High Performance Computing
  Applications}, 36\penalty0 (1):\penalty0 106--128, 2022.

\bibitem[Mehl et~al.(2016)Mehl, Uekermann, Bijl, Blom, Gatzhammer, and van
  Zuijlen]{mehlParallelCouplingNumerics2016}
Miriam Mehl, Benjamin Uekermann, Hester Bijl, David Blom, Bernhard Gatzhammer,
  and Alexander van Zuijlen.
\newblock Parallel coupling numerics for partitioned fluid–structure
  interaction simulations.
\newblock \emph{Computers \& Mathematics with Applications}, 71\penalty0
  (4):\penalty0 869--891, February 2016.
\newblock ISSN 08981221.
\newblock \doi{10.1016/j.camwa.2015.12.025}.
\newblock URL
  \url{https://linkinghub.elsevier.com/retrieve/pii/S0898122115005933}.

\bibitem[Parker et~al.(2006)Parker, Guilkey, and
  Harman]{parkerComponentbasedParallelInfrastructure2006}
Steven~G Parker, James Guilkey, and Todd Harman.
\newblock A {Component}-based {Parallel} {Infrastructure} for the {Simulation}
  of {Fluid}-structure {Interaction}.
\newblock \emph{Eng. with Comput.}, 22\penalty0 (3-4):\penalty0 277--292, 2006.

\bibitem[Cheng et~al.(2020)Cheng, Dominski, Chen, Chen, Merlo, Ku, Hager,
  Chang, Suchyta, D'azevedo, Ethier, Sreepathi, Klasky, Jenko, Bhattacharjee,
  and Parker]{chengSpatialCoreedgeCoupling2020}
Junyi Cheng, Julien Dominski, Yang Chen, Haotian Chen, Gabriele Merlo,
  Seung-Hoe Ku, Robert Hager, C.S. Chang, Eric Suchyta, Ed~D'azevedo, Stephane
  Ethier, Sarat Sreepathi, Scott Klasky, Frank Jenko, Amitava Bhattacharjee,
  and Scott Parker.
\newblock Spatial core-edge coupling of the {PIC} gyrokinetic codes {GEM} and
  {XGC}.
\newblock \emph{Physics of Plasmas}, 27\penalty0 (12):\penalty0 122510,
  December 2020.
\newblock \doi{10.1063/5.0026043}.

\bibitem[Jo et~al.(2025)Jo, Seo, Kwon, and
  Yoon]{joFieldalignedGyrokineticSolver2025}
Gahyung Jo, Janghoon Seo, Jae-Min Kwon, and Eisung Yoon.
\newblock A field-aligned gyrokinetic solver based on discontinuous galerkin in
  tokamak geometry.
\newblock \emph{Computer Physics Communications}, 316:\penalty0 109769,
  November 2025.
\newblock ISSN 0010-4655.
\newblock \doi{10.1016/j.cpc.2025.109769}.

\bibitem[Alauzet(2016)]{alauzetParallelMatrixfreeConservative2016}
Frédéric Alauzet.
\newblock A parallel matrix-free conservative solution interpolation on
  unstructured tetrahedral meshes.
\newblock \emph{Computer Methods in Applied Mechanics and Engineering},
  299:\penalty0 116--142, February 2016.
\newblock ISSN 00457825.
\newblock \doi{10.1016/j.cma.2015.10.012}.
\newblock URL
  \url{https://linkinghub.elsevier.com/retrieve/pii/S004578251500331X}.

\bibitem[Slattery(2016)]{slatteryMeshfreeDataTransfer2016}
Stuart~R. Slattery.
\newblock Mesh-free data transfer algorithms for partitioned multiphysics
  problems: {{Conservation}}, accuracy, and parallelism.
\newblock \emph{Journal of Computational Physics}, 307:\penalty0 164--188,
  2016.

\bibitem[Bungartz et~al.(2016)Bungartz, Lindner, Gatzhammer, Mehl, Scheufele,
  Shukaev, and Uekermann]{bungartzPreCICEFullyParallel2016}
Hans-Joachim Bungartz, Florian Lindner, Bernhard Gatzhammer, Miriam Mehl,
  Klaudius Scheufele, Alexander Shukaev, and Benjamin Uekermann.
\newblock {preCICE} – {A} fully parallel library for multi-physics surface
  coupling.
\newblock \emph{Computers \& Fluids}, 141:\penalty0 250--258, December 2016.
\newblock ISSN 00457930.
\newblock \doi{10.1016/j.compfluid.2016.04.003}.
\newblock URL
  \url{https://linkinghub.elsevier.com/retrieve/pii/S0045793016300974}.

\bibitem[Mollén et~al.(2021)Mollén, Adams, Knepley, Hager, and
  Chang]{mollenImplementationHigherorderVelocity2021}
Albert Mollén, M.~F. Adams, M.~G. Knepley, R.~Hager, and C.~S. Chang.
\newblock Implementation of higher-order velocity mapping between marker
  particles and grid in the particle-in-cell code {{XGC}}.
\newblock \emph{Journal of Plasma Physics}, 87\penalty0 (2):\penalty0
  905870229, 2021.

\bibitem[Jiao and Heath(2004)]{jiaoCommonrefinementbasedDataTransfer2004}
Xiangmin Jiao and Michael~T. Heath.
\newblock Common-refinement-based data transfer between non-matching meshes in
  multiphysics simulations.
\newblock \emph{International Journal for Numerical Methods in Engineering},
  61\penalty0 (14):\penalty0 2402--2427, 2004.

\bibitem[Blanchard and LoubËre(2016)]{blanchardHighOrderAccurate2016}
Ghislain Blanchard and RaphaÎl LoubËre.
\newblock High order accurate conservative remapping scheme on polygonal meshes
  using a posteriori {{MOOD}} limiting.
\newblock \emph{Computers \& Fluids}, 136:\penalty0 83--103, 2016.

\bibitem[Jaiman et~al.(2006)Jaiman, Jiao, Geubelle, and
  Loth]{jaimanConservativeLoadTransfer2006}
R.K. Jaiman, X.~Jiao, P.H. Geubelle, and E.~Loth.
\newblock Conservative load transfer along curved fluid–solid interface with
  non-matching meshes.
\newblock \emph{Journal of Computational Physics}, 218\penalty0 (1):\penalty0
  372--397, October 2006.
\newblock ISSN 00219991.
\newblock \doi{10.1016/j.jcp.2006.02.016}.
\newblock URL
  \url{https://linkinghub.elsevier.com/retrieve/pii/S0021999106000891}.

\bibitem[Menon and
  Schmidt(2011)]{menonConservativeInterpolationUnstructured2011}
Sandeep Menon and David~P. Schmidt.
\newblock Conservative interpolation on unstructured polyhedral meshes: {An}
  extension of the supermesh approach to cell-centered finite-volume variables.
\newblock \emph{Computer Methods in Applied Mechanics and Engineering},
  200\penalty0 (41-44):\penalty0 2797--2804, October 2011.
\newblock ISSN 00457825.
\newblock \doi{10.1016/j.cma.2011.04.025}.
\newblock URL
  \url{https://linkinghub.elsevier.com/retrieve/pii/S0045782511001666}.

\bibitem[Farrell and
  Maddison(2011)]{farrellConservativeInterpolationVolume2011}
P.E. Farrell and J.R. Maddison.
\newblock Conservative interpolation between volume meshes by local
  {{Galerkin}} projection.
\newblock \emph{Computer Methods in Applied Mechanics and Engineering},
  200\penalty0 (1-4):\penalty0 89--100, 2011.

\bibitem[Farrell et~al.(2009)Farrell, Piggott, Pain, Gorman, and
  Wilson]{farrellConservativeInterpolationUnstructured2009}
P.E. Farrell, M.D. Piggott, C.C. Pain, G.J. Gorman, and C.R. Wilson.
\newblock Conservative interpolation between unstructured meshes via supermesh
  construction.
\newblock \emph{Computer Methods in Applied Mechanics and Engineering},
  198\penalty0 (33-36):\penalty0 2632--2642, July 2009.
\newblock ISSN 00457825.
\newblock \doi{10.1016/j.cma.2009.03.004}.
\newblock URL
  \url{https://linkinghub.elsevier.com/retrieve/pii/S0045782509001315}.

\bibitem[Edwards et~al.(2014)Edwards, Trott, and Sunderland]{kokkos2014}
H.~Carter Edwards, Christian~R. Trott, and Daniel Sunderland.
\newblock Kokkos: Enabling manycore performance portability through polymorphic
  memory access patterns.
\newblock \emph{Journal of Parallel and Distributed Computing}, 74\penalty0
  (12):\penalty0 3202 -- 3216, 2014.
\newblock ISSN 0743-7315.
\newblock \doi{https://doi.org/10.1016/j.jpdc.2014.07.003}.
\newblock URL
  \url{http://www.sciencedirect.com/science/article/pii/S0743731514001257}.

\bibitem[Zienkiewicz and
  Zhu(1992{\natexlab{a}})]{zienkiewiczSuperconvergentPatchRecovery1992}
O.C. Zienkiewicz and J.Z. Zhu.
\newblock The superconvergent patch recovery (spr) and adaptive finite element
  refinement.
\newblock \emph{Computer Methods in Applied Mechanics and Engineering},
  101\penalty0 (1-3):\penalty0 207--224, December 1992{\natexlab{a}}.
\newblock ISSN 00457825.
\newblock \doi{10.1016/0045-7825(92)90023-D}.

\bibitem[Zienkiewicz and
  Zhu(1992{\natexlab{b}})]{zienkiewiczSuperconvergentPatchRecovery1992b}
O.~C. Zienkiewicz and J.~Z. Zhu.
\newblock The superconvergent patch recovery anda posteriori error estimates.
  part 2: Error estimates and adaptivity.
\newblock \emph{International Journal for Numerical Methods in Engineering},
  33\penalty0 (7):\penalty0 1365--1382, May 1992{\natexlab{b}}.
\newblock ISSN 0029-5981, 1097-0207.
\newblock \doi{10.1002/nme.1620330703}.

\bibitem[Riaz(2024)]{riaz_automated_2024}
Usman Riaz.
\newblock \emph{An {Automated} {Modeling}, {Meshing}, and {Adaptive}
  {Framework} for {Tokamak} {Plasma} {Simulations}}.
\newblock Ph.{D}., Rensselaer Polytechnic Institute, United States -- New York,
  2024.
\newblock URL
  \url{https://www.proquest.com/docview/3071391606/abstract/56D705E116C146AAPQ/1}.
\newblock ISBN: 9798383059623.

\bibitem[Powell(2015)]{powellR3dSoftwareFast2015}
Devon Powell.
\newblock r3d: {Software} for fast, robust geometric operations in {3D} and
  {2D}.
\newblock Technical Report LA-UR-15-26964, August 2015.

\bibitem[Ibanez(2016)]{ibanez2016conformal}
Daniel~Alejandro Ibanez.
\newblock Conformal mesh adaptation on heterogeneous supercomputers, 2016.
\newblock {{Ph.D. thesis}}.

\bibitem[Jackins and Tanimoto(1980)]{jackinsOcttreesTheirUse1980}
Chris~L. Jackins and Steven~L. Tanimoto.
\newblock Oct-trees and their use in representing three-dimensional objects.
\newblock \emph{Computer Graphics and Image Processing}, 14\penalty0
  (3):\penalty0 249--270, November 1980.
\newblock ISSN 0146-664X.
\newblock \doi{10.1016/0146-664X(80)90055-6}.
\newblock URL
  \url{https://www.sciencedirect.com/science/article/pii/0146664X80900556}.

\bibitem[Popov et~al.(2007)Popov, Günther, Seidel, and
  Slusallek]{popovStacklessKDTreeTraversal2007}
Stefan Popov, Johannes Günther, Hans-Peter Seidel, and Philipp Slusallek.
\newblock Stackless {KD}-{Tree} {Traversal} for {High} {Performance} {GPU}
  {Ray} {Tracing}.
\newblock \emph{Computer Graphics Forum}, 26\penalty0 (3):\penalty0 415--424,
  2007.
\newblock ISSN 1467-8659.
\newblock \doi{10.1111/j.1467-8659.2007.01064.x}.
\newblock URL
  \url{https://onlinelibrary.wiley.com/doi/abs/10.1111/j.1467-8659.2007.01064.x}.
\newblock \_eprint:
  https://onlinelibrary.wiley.com/doi/pdf/10.1111/j.1467-8659.2007.01064.x.

\bibitem[Beckmann et~al.(1990)Beckmann, Kriegel, Schneider, and
  Seeger]{beckmannRtreeEfficientRobust1990}
Norbert Beckmann, Hans-Peter Kriegel, Ralf Schneider, and Bernhard Seeger.
\newblock The {R}*-tree: an efficient and robust access method for points and
  rectangles.
\newblock In \emph{Proceedings of the 1990 {ACM} {SIGMOD} international
  conference on {Management} of data}, {SIGMOD} '90, pages 322--331, New York,
  NY, USA, May 1990. Association for Computing Machinery.
\newblock ISBN 978-0-89791-365-2.
\newblock \doi{10.1145/93597.98741}.
\newblock URL \url{https://dl.acm.org/doi/10.1145/93597.98741}.

\bibitem[Lebrun-Grandié et~al.(2020)Lebrun-Grandié, Prokopenko, Turcksin, and
  Slattery]{lebrun-grandieArborXPerformancePortable2020}
D.~Lebrun-Grandié, A.~Prokopenko, B.~Turcksin, and S.~R. Slattery.
\newblock {ArborX}: {A} {Performance} {Portable} {Geometric} {Search}
  {Library}.
\newblock \emph{ACM Transactions on Mathematical Software}, 47\penalty0
  (1):\penalty0 1--15, December 2020.
\newblock ISSN 0098-3500, 1557-7295.
\newblock \doi{10.1145/3412558}.
\newblock URL \url{https://dl.acm.org/doi/10.1145/3412558}.

\bibitem[Lauterbach et~al.(2009)Lauterbach, Garland, Sengupta, Luebke, and
  Manocha]{lauterbachFastBVHConstruction2009}
C.~Lauterbach, M.~Garland, S.~Sengupta, D.~Luebke, and D.~Manocha.
\newblock Fast {BVH} {Construction} on {GPUs}.
\newblock \emph{Computer Graphics Forum}, 28\penalty0 (2):\penalty0 375--384,
  2009.
\newblock ISSN 1467-8659.
\newblock \doi{10.1111/j.1467-8659.2009.01377.x}.
\newblock URL
  \url{https://onlinelibrary.wiley.com/doi/abs/10.1111/j.1467-8659.2009.01377.x}.
\newblock \_eprint:
  https://onlinelibrary.wiley.com/doi/pdf/10.1111/j.1467-8659.2009.01377.x.

\bibitem[Foley and Sugerman(2005)]{foleyKDtreeAccelerationStructures2005}
Tim Foley and Jeremy Sugerman.
\newblock {KD}-tree acceleration structures for a {GPU} raytracer.
\newblock In \emph{Proceedings of the {ACM} {SIGGRAPH}/{EUROGRAPHICS}
  conference on {Graphics} hardware - {HWWS} '05}, page~15, Los Angeles,
  California, 2005. ACM Press.
\newblock ISBN 978-1-59593-086-6.
\newblock \doi{10.1145/1071866.1071869}.
\newblock URL \url{http://portal.acm.org/citation.cfm?doid=1071866.1071869}.

\bibitem[Morrical et~al.(2022)Morrical, Wald, Usher, and
  Pascucci]{morricalAcceleratingUnstructuredMesh2022}
Nate Morrical, Ingo Wald, Will Usher, and Valerio Pascucci.
\newblock Accelerating {Unstructured} {Mesh} {Point} {Location} {With} {RT}
  {Cores}.
\newblock \emph{IEEE Transactions on Visualization and Computer Graphics},
  28\penalty0 (8):\penalty0 2852--2866, August 2022.
\newblock ISSN 1077-2626, 1941-0506, 2160-9306.
\newblock \doi{10.1109/TVCG.2020.3042930}.
\newblock URL \url{https://ieeexplore.ieee.org/document/9286513/}.

\bibitem[Menezes et~al.(2022)Menezes, De~Magalhães, De~Oliveira,
  Randolph~Franklin, and Coelho]{menezesEmployingGPUsAccelerate2022}
Marcelo Menezes, Salles V.~G. De~Magalhães, Matheus~Aguilar De~Oliveira,
  W.~Randolph~Franklin, and Bruno Coelho.
\newblock Employing {GPUs} to {Accelerate} {Exact} {Geometric} {Predicates} for
  {3D} {Geospatial} {Processing}.
\newblock In John Krumm, Andreas Züfle, and Cyrus Shahabi, editors,
  \emph{Spatial {Gems}, {Volume} 1}, pages 97--110. ACM, New York, NY, USA, 1
  edition, August 2022.
\newblock ISBN 978-1-4503-9813-8.
\newblock \doi{10.1145/3548732.3548744}.
\newblock URL \url{https://dl.acm.org/doi/10.1145/3548732.3548744}.

\bibitem[Akman et~al.(1989)Akman, Franklin, Kankanhalli, and
  Narayanaswami]{akmanGeometricComputingUniform1989}
V.~Akman, W.R. Franklin, M.~Kankanhalli, and C.~Narayanaswami.
\newblock Geometric computing and uniform grid technique.
\newblock \emph{Computer-Aided Design}, 21\penalty0 (7):\penalty0 410--420,
  September 1989.
\newblock ISSN 00104485.
\newblock \doi{10.1016/0010-4485(89)90125-5}.
\newblock URL
  \url{https://linkinghub.elsevier.com/retrieve/pii/0010448589901255}.

\bibitem[Zlatuška and Havran(2010)]{zlatuskaRayTracingGpu2010}
Martin Zlatuška and Vlastimil Havran.
\newblock Ray tracing on a gpu with cuda–comparative study of three
  algorithms.
\newblock In \emph{Proceedings of {WSCG}}, pages 69--75, 2010.
\newblock URL
  \url{https://www.researchgate.net/profile/Vlastimil-Havran/publication/228974703_Ray_Tracing_on_a_GPU_with_CUDA-comparative_study_of_three_algorithms/links/09e4150fedd72ad8d5000000/Ray-Tracing-on-a-GPU-with-CUDA-comparative-study-of-three-algorithms.pdf}.

\bibitem[Lubbe et~al.(2020)Lubbe, Xu, Wilke, Pizette, and
  Govender]{lubbeAnalysisParallelSpatial2020}
Retief Lubbe, Wen-Jie Xu, Daniel~N. Wilke, Patrick Pizette, and Nicolin
  Govender.
\newblock Analysis of parallel spatial partitioning algorithms for {GPU} based
  {DEM}.
\newblock \emph{Computers and Geotechnics}, 125:\penalty0 103708, September
  2020.
\newblock ISSN 0266352X.
\newblock \doi{10.1016/j.compgeo.2020.103708}.
\newblock URL
  \url{https://linkinghub.elsevier.com/retrieve/pii/S0266352X20302718}.

\bibitem[Ibanez et~al.(2016)Ibanez, Seol, Smith, and Shephard]{ibanez2016pumi}
Daniel~A Ibanez, E~Seegyoung Seol, Cameron~W Smith, and Mark~S Shephard.
\newblock {PUMI:} parallel unstructured mesh infrastructure.
\newblock \emph{ACM Transactions on Mathematical Software (TOMS)}, 42\penalty0
  (3):\penalty0 1--28, 2016.

\bibitem[Riaz et~al.(2024)Riaz, Seol, Hager, and Shephard]{riaz2023modeling}
Usman Riaz, E~Seegyoung Seol, Robert Hager, and Mark~S Shephard.
\newblock Modeling and meshing for tokamak edge plasma simulations.
\newblock \emph{Computer Physics Communications}, 295:\penalty0 108982, 2024.

\bibitem[Merson and Shephard(2021)]{mersonModeltraitsModelAttribute2021}
Jacob Merson and Mark~S Shephard.
\newblock Model-traits: Model attribute definitions for scientific simulations
  in c++.
\newblock \emph{The Journal of Open Source Software}, 6\penalty0 (64):\penalty0
  3389, August 2021.
\newblock \doi{10.21105/joss.03389}.

\bibitem[Diamond et~al.(2021)Diamond, Smith, Zhang, Yoon, and
  Shephard]{diamond2021pumipic}
Gerrett Diamond, Cameron~W Smith, Chonglin Zhang, Eisung Yoon, and Mark~S
  Shephard.
\newblock {PUMIPic}: A mesh-based approach to unstructured mesh
  particle-in-cell on {GPUs}.
\newblock \emph{Journal of Parallel and Distributed Computing}, 157:\penalty0
  1--12, 2021.

\bibitem[Hasan et~al.(2025)Hasan, Smith, Shephard, Churchill, Wilkie, Romano,
  Shriwise, and Merson]{hasanGPUAccelerationMonte2025}
Fuad Hasan, Cameron~W. Smith, Mark~S. Shephard, R.~Michael Churchill, George~J.
  Wilkie, Paul~K. Romano, Patrick~C. Shriwise, and Jacob~S. Merson.
\newblock {GPU} {Acceleration} of {Monte} {Carlo} {Tallies} on {Unstructured}
  {Meshes} in {OpenMC} with {PUMI}-{Tally}, April 2025.
\newblock URL \url{http://arxiv.org/abs/2504.19048}.
\newblock arXiv:2504.19048 [cs].

\bibitem[D’haeseleer et~al.(1991)D’haeseleer, Hitchon, Callen, and
  Shohet]{dhaeseleerFluxCoordinatesMagnetic1991}
William~Denis D’haeseleer, William Nicholas~Guy Hitchon, James~D. Callen, and
  J.~Leon Shohet.
\newblock \emph{Flux {Coordinates} and {Magnetic} {Field} {Structure}}.
\newblock Springer Berlin Heidelberg, Berlin, Heidelberg, 1991.
\newblock ISBN 978-3-642-75597-2 978-3-642-75595-8.
\newblock \doi{10.1007/978-3-642-75595-8}.
\newblock URL \url{http://link.springer.com/10.1007/978-3-642-75595-8}.

\bibitem[Boozer(1981)]{boozerPlasmaEquilibriumRational1981}
Allen~H Boozer.
\newblock Plasma equilibrium with rational magnetic surfaces.
\newblock Technical report, Princeton Plasma Physics Lab. (PPPL), Princeton, NJ
  (United States), 1981.
\newblock URL \url{https://www.osti.gov/servlets/purl/6861751-q8CuSt/}.

\bibitem[Kruger and Greene(2019)]{krugerRelationshipFluxCoordinates2019}
S.~E. Kruger and John~M. Greene.
\newblock The relationship between flux coordinates and equilibrium-based
  frames of reference in fusion theory.
\newblock \emph{Physics of Plasmas}, 26\penalty0 (8):\penalty0 082506, August
  2019.
\newblock ISSN 1070-664X.
\newblock \doi{10.1063/1.5098313}.
\newblock URL \url{https://doi.org/10.1063/1.5098313}.

\bibitem[Lin et~al.(1998)Lin, Hahm, Lee, Tang, and
  White]{linTurbulentTransportReduction1998}
Z.~Lin, T.~S. Hahm, W.~W. Lee, W.~M. Tang, and R.~B. White.
\newblock Turbulent {Transport} {Reduction} by {Zonal} {Flows}: {Massively}
  {Parallel} {Simulations}.
\newblock \emph{Science}, 281\penalty0 (5384):\penalty0 1835--1837, September
  1998.
\newblock ISSN 0036-8075, 1095-9203.
\newblock \doi{10.1126/science.281.5384.1835}.
\newblock URL \url{https://www.science.org/doi/10.1126/science.281.5384.1835}.

\bibitem[Hirshman and Whitson(1983)]{hirshmanSTEEPESTDESCENTMOMENT}
S~P Hirshman and J~C Whitson.
\newblock Steepest descent moment method for three-dimensional
  magnetohydrodynamic equilibria.
\newblock 1983.

\bibitem[McMillan and Lazerson(2014)]{mcmillanBEAMS3DNeutralBeam2014}
Matthew McMillan and Samuel~A Lazerson.
\newblock {BEAMS3D} {Neutral} {Beam} {Injection} {Model}*.
\newblock \emph{Plasma Physics and Controlled Fusion}, 56\penalty0
  (9):\penalty0 095019, July 2014.
\newblock ISSN 0741-3335.
\newblock \doi{10.1088/0741-3335/56/9/095019}.
\newblock URL \url{https://doi.org/10.1088/0741-3335/56/9/095019}.
\newblock Publisher: IOP Publishing.

\bibitem[Jenko et~al.(2000)Jenko, Dorland, Kotschenreuther, and
  Rogers]{jenkoElectronTemperatureGradient2000}
Frank Jenko, William Dorland, M.~Kotschenreuther, and B.~Rogers.
\newblock Electron temperature gradient driven turbulence.
\newblock \emph{Physics of Plasmas - PHYS PLASMAS}, 7:\penalty0 1904--1910, May
  2000.
\newblock \doi{10.1063/1.874014}.

\bibitem[Hariri and
  Ottaviani(2013)]{haririFluxcoordinateIndependentFieldaligned2013}
F.~Hariri and M.~Ottaviani.
\newblock A flux-coordinate independent field-aligned approach to plasma
  turbulence simulations.
\newblock \emph{Computer Physics Communications}, 184\penalty0 (11):\penalty0
  2419--2429, November 2013.
\newblock ISSN 00104655.
\newblock \doi{10.1016/j.cpc.2013.06.005}.
\newblock URL
  \url{https://linkinghub.elsevier.com/retrieve/pii/S0010465513001999}.

\bibitem[Stegmeir et~al.(2017)Stegmeir, Maj, Coster, Lackner, Held, and
  Wiesenberger]{stegmeirAdvancesFluxcoordinateIndependent2017}
Andreas Stegmeir, Omar Maj, David Coster, Karl Lackner, Markus Held, and
  Matthias Wiesenberger.
\newblock Advances in the flux-coordinate independent approach.
\newblock \emph{Computer Physics Communications}, 213:\penalty0 111--121, April
  2017.
\newblock ISSN 0010-4655.
\newblock \doi{10.1016/j.cpc.2016.12.014}.
\newblock URL
  \url{https://www.sciencedirect.com/science/article/pii/S0010465516303939}.

\bibitem[Shanahan et~al.(2018)Shanahan, Dudson, and
  Hill]{shanahanFluidSimulationsPlasma2018}
B.~Shanahan, B.~Dudson, and P.~Hill.
\newblock Fluid simulations of plasma filaments in stellarator geometries with
  {BSTING}.
\newblock \emph{Plasma Physics and Controlled Fusion}, 61\penalty0
  (2):\penalty0 025007, December 2018.
\newblock ISSN 0741-3335.
\newblock \doi{10.1088/1361-6587/aaed7d}.
\newblock URL \url{https://dx.doi.org/10.1088/1361-6587/aaed7d}.
\newblock Publisher: IOP Publishing.

\bibitem[Michels et~al.(2021)Michels, Stegmeir, Ulbl, Jarema, and
  Jenko]{michelsGENEXFullfGyrokinetic2021}
Dominik Michels, Andreas Stegmeir, Philipp Ulbl, Denis Jarema, and Frank Jenko.
\newblock {GENE}-{X}: {A} full-\textit{f} gyrokinetic turbulence code based on
  the flux-coordinate independent approach.
\newblock \emph{Computer Physics Communications}, 264:\penalty0 107986, July
  2021.
\newblock ISSN 0010-4655.
\newblock \doi{10.1016/j.cpc.2021.107986}.
\newblock URL
  \url{https://www.sciencedirect.com/science/article/pii/S0010465521000989}.

\bibitem[Ku et~al.(2016)Ku, Hager, Chang, Kwon, and
  Parker]{kuNewHybridLagrangianNumerical2016}
S.~Ku, R.~Hager, C.~S. Chang, J.~M. Kwon, and S.~E. Parker.
\newblock A new hybrid-{Lagrangian} numerical scheme for gyrokinetic simulation
  of tokamak edge plasma.
\newblock \emph{Journal of Computational Physics}, 315:\penalty0 467--475, June
  2016.
\newblock ISSN 0021-9991.
\newblock \doi{10.1016/j.jcp.2016.03.062}.
\newblock URL
  \url{https://www.sciencedirect.com/science/article/pii/S0021999116300274}.

\bibitem[Moritaka et~al.(2019)Moritaka, Hager, Cole, Lazerson, Chang, Ku,
  Matsuoka, Satake, and
  Ishiguro]{moritakaDevelopmentGyrokineticParticleCell2019}
Toseo Moritaka, Robert Hager, Michael Cole, Samuel Lazerson, Choong-Seock
  Chang, Seung-Hoe Ku, Seikichi Matsuoka, Shinsuke Satake, and Seiji Ishiguro.
\newblock Development of a {Gyrokinetic} {Particle}-in-{Cell} {Code} for
  {Whole}-{Volume} {Modeling} of {Stellarators}.
\newblock \emph{Plasma}, 2\penalty0 (2):\penalty0 179--200, June 2019.
\newblock ISSN 2571-6182.
\newblock \doi{10.3390/plasma2020014}.
\newblock URL \url{https://www.mdpi.com/2571-6182/2/2/14}.
\newblock Number: 2 Publisher: Multidisciplinary Digital Publishing Institute.

\bibitem[Sovinec et~al.(2004)Sovinec, Glasser, Gianakon, Barnes, Nebel, Kruger,
  Schnack, Plimpton, Tarditi, and
  Chu]{sovinecNonlinearMagnetohydrodynamicsSimulation2004}
C.R. Sovinec, A.H. Glasser, T.A. Gianakon, D.C. Barnes, R.A. Nebel, S.E.
  Kruger, D.D. Schnack, S.J. Plimpton, A.~Tarditi, and M.S. Chu.
\newblock Nonlinear magnetohydrodynamics simulation using high-order finite
  elements.
\newblock \emph{Journal of Computational Physics}, 195\penalty0 (1):\penalty0
  355--386, March 2004.
\newblock ISSN 00219991.
\newblock \doi{10.1016/j.jcp.2003.10.004}.
\newblock URL
  \url{https://linkinghub.elsevier.com/retrieve/pii/S0021999103005369}.

\bibitem[Candy et~al.(2016)Candy, Belli, and
  Bravenec]{candyHighaccuracyEulerianGyrokinetic2016}
J.~Candy, E.~A. Belli, and R.~V. Bravenec.
\newblock A high-accuracy {Eulerian} gyrokinetic solver for collisional
  plasmas.
\newblock \emph{Journal of Computational Physics}, 324:\penalty0 73--93,
  November 2016.
\newblock ISSN 0021-9991.
\newblock \doi{10.1016/j.jcp.2016.07.039}.
\newblock URL
  \url{https://www.sciencedirect.com/science/article/pii/S0021999116303400}.

\bibitem[Plimpton and Knight(2021)]{plimptonRendezvousAlgorithmsLargescale2021}
Steven~J. Plimpton and Christopher Knight.
\newblock Rendezvous algorithms for large-scale modeling and simulation.
\newblock \emph{Journal of Parallel and Distributed Computing}, 147:\penalty0
  184--195, January 2021.
\newblock ISSN 0743-7315.
\newblock \doi{10.1016/j.jpdc.2020.09.001}.

\bibitem[Sebastian~Rettenberger(2014)]{rettenberger14}
Christian~Pelties Sebastian~Rettenberger, C.W.~Smith.
\newblock Optimizing {CAD} and mesh generation workflow for {SeisSol}.
\newblock In \emph{Proceedings of the International Conference for High
  Performance Computing, Networking, Storage and Analysis}, New Orleans, LA,
  November 2014.

\bibitem[BEALL and SHEPHARD(1997)]{beallShephardMeshDataStructure}
MARK~W. BEALL and MARK~S. SHEPHARD.
\newblock A general topology-based mesh data structure.
\newblock \emph{International Journal for Numerical Methods in Engineering},
  40\penalty0 (9):\penalty0 1573--1596, 1997.
\newblock
  \doi{https://doi.org/10.1002/(SICI)1097-0207(19970515)40:9<1573::AID-NME128>3.0.CO;2-9}.
\newblock URL
  \url{https://onlinelibrary.wiley.com/doi/abs/10.1002/%28SICI%291097-0207%2819970515%2940%3A9%3C1573%3A%3AAID-NME128%3E3.0.CO%3B2-9}.

\bibitem[Docan et~al.(2010)Docan, Parashar, and
  Klasky]{docanDataSpacesInteractionCoordination2010}
Ciprian Docan, Manish Parashar, and Scott Klasky.
\newblock {DataSpaces}: an interaction and coordination framework for coupled
  simulation workflows.
\newblock In \emph{Proceedings of the 19th {ACM} {International} {Symposium} on
  {High} {Performance} {Distributed} {Computing}}, pages 25--36, Chicago
  Illinois, June 2010. ACM.
\newblock ISBN 978-1-60558-942-8.
\newblock \doi{10.1145/1851476.1851481}.
\newblock URL \url{https://dl.acm.org/doi/10.1145/1851476.1851481}.

\bibitem[Davis et~al.(2023)Davis, Merson, Subedi, Ricketson, Smith, Shephard,
  and Parashar]{davis2023benesh}
P.~Davis, J.~Merson, P.~Subedi, L.~Ricketson, C.~Smith, M.S. Shephard, and
  M.~Parashar.
\newblock Benesh: Choreographic coordination for in-situ workflows.
\newblock In \emph{30th IEEE International Conference on High Performance
  Computing, Data, and Analytics (HiPC)}. IEEE, 2023.

\bibitem[Zhang et~al.(2023)Zhang, Diamond, Smith, and
  Shephard]{zhang2023development}
Chonglin Zhang, Gerrett Diamond, Cameron~W Smith, and Mark~S Shephard.
\newblock Development of an unstructured mesh gyrokinetic particle-in-cell code
  for exascale fusion plasma simulations on {GPUs}.
\newblock \emph{Computer Physics Communications}, page 108824, 2023.

\bibitem[Murakami et~al.(2000)Murakami, Gasparino, Idei, Kubo, Maassberg,
  Marushchenko, Nakajima, Romé, and Okamoto]{murakami5DSimulationStudy2000}
S~Murakami, U~Gasparino, H~Idei, S~Kubo, H~Maassberg, N~Marushchenko,
  N~Nakajima, M~Romé, and M~Okamoto.
\newblock 5-{D} simulation study of suprathermal electron transport in
  non-axisymmetric plasmas.
\newblock \emph{Nuclear Fusion}, 40\penalty0 (3Y):\penalty0 693--700, March
  2000.
\newblock ISSN 0029-5515.
\newblock \doi{10.1088/0029-5515/40/3Y/333}.
\newblock URL
  \url{https://iopscience.iop.org/article/10.1088/0029-5515/40/3Y/333}.

\bibitem[Williams(1991)]{williamsRIB}
Roy~D. Williams.
\newblock Performance of dynamic load balancing algorithms for unstructured
  mesh calculations.
\newblock \emph{Concurrency: Practice and Experience}, 3\penalty0 (5):\penalty0
  457--481, 1991.
\newblock \doi{https://doi.org/10.1002/cpe.4330030502}.
\newblock URL
  \url{https://onlinelibrary.wiley.com/doi/abs/10.1002/cpe.4330030502}.

\bibitem[Meneghini et~al.(2021)Meneghini, Snoep, Lyons, McClenaghan, Imai,
  Grierson, Smith, Staebler, Snyder, Candy, Belli, Lao, Park, Citrin,
  Cordemiglia, Tema, and
  Mordijck]{meneghiniNeuralnetworkAcceleratedCoupled2021}
O.~Meneghini, G.~Snoep, B.C. Lyons, J.~McClenaghan, C.S. Imai, B.~Grierson,
  S.P. Smith, G.M. Staebler, P.B. Snyder, J.~Candy, E.~Belli, L.~Lao, J.M.
  Park, J.~Citrin, T.L. Cordemiglia, A.~Tema, and S.~Mordijck.
\newblock Neural-network accelerated coupled core-pedestal simulations with
  self-consistent transport of impurities and compatible with {ITER} {IMAS}.
\newblock \emph{Nuclear Fusion}, 61\penalty0 (2):\penalty0 026006, February
  2021.
\newblock ISSN 0029-5515, 1741-4326.
\newblock \doi{10.1088/1741-4326/abb918}.
\newblock URL
  \url{https://iopscience.iop.org/article/10.1088/1741-4326/abb918}.

\bibitem[Hindmarsh et~al.(2005)Hindmarsh, Brown, Grant, Lee, Serban, Shumaker,
  and Woodward]{hindmarsh2005sundials}
Alan~C Hindmarsh, Peter~N Brown, Keith~E Grant, Steven~L Lee, Radu Serban,
  Dan~E Shumaker, and Carol~S Woodward.
\newblock {SUNDIALS}: Suite of nonlinear and differential/algebraic equation
  solvers.
\newblock \emph{ACM Transactions on Mathematical Software (TOMS)}, 31\penalty0
  (3):\penalty0 363--396, 2005.
\newblock \doi{10.1145/1089014.1089020}.

\bibitem[Gardner et~al.(2022)Gardner, Reynolds, Woodward, and
  Balos]{gardner2022sundials}
David~J Gardner, Daniel~R Reynolds, Carol~S Woodward, and Cody~J Balos.
\newblock Enabling new flexibility in the {SUNDIALS} suite of nonlinear and
  differential/algebraic equation solvers.
\newblock \emph{ACM Transactions on Mathematical Software (TOMS)}, 48\penalty0
  (3):\penalty0 1--24, 2022.
\newblock \doi{10.1145/3539801}.

\end{thebibliography}

\end{document}